\documentclass[manuscript,screen]{acmart}

\usepackage{multirow}
\usepackage{booktabs}
\usepackage{pifont}
\usepackage{enumitem}
\usepackage{makecell}
\usepackage{algorithm}
\usepackage{listings}
\usepackage{algorithmic}
\usepackage{amsthm,amsmath}
\usepackage{subfigure}
\usepackage[misc]{ifsym}
\usepackage{xcolor}
\usepackage{graphicx}
\usepackage{stfloats}
\usepackage{float}
\usepackage{url}
\usepackage{color}

\newcommand{\en}[1]{{\color{black}{#1}}}

\AtBeginDocument{%
  }

\setcopyright{acmlicensed}
\copyrightyear{2018}
\acmYear{2018}
\acmDOI{XXXXXXX.XXXXXXX}

\acmConference[Conference acronym 'XX]{Make sure to enter the correct
  conference title from your rights confirmation emai}{June 03--05,
  2018}{Woodstock, NY}
\acmISBN{978-1-4503-XXXX-X/18/06}

\begin{document}


\title{Repeated Padding+: Simple yet Effective Data Augmentation Plugin for Sequential Recommendation}



\author{Yizhou Dang}
\affiliation{%
  \institution{Software College, Northeastern University}
  \city{Shenyang}
  \country{China}
}
\email{dangyz@stumail.neu.edu.cn}

\author{Yuting Liu}
\affiliation{%
  \institution{Software College, Northeastern University}
  \city{Shenyang}
  \country{China}
}
\email{yutingliu@stumail.neu.edu.cn}

\author{Enneng Yang}
\affiliation{%
  \institution{Software College, Northeastern University}
  \city{Shenyang}
  \country{China}
}
\email{ennengyang@stumail.neu.edu.cn}

\author{Guibing Guo}
\authornote{Corresponding authors.}
\affiliation{%
  \institution{Software College, Northeastern University}
  \city{Shenyang}
  \country{China}
}
\email{guogb@swc.neu.edu.cn}

\author{Linying	Jiang}
\affiliation{%
  \institution{Software College, Northeastern University}
  \city{Shenyang}
  \country{China}
}
\email{jiangly@swc.neu.edu.cn}

\author{Jianzhe Zhao}
\affiliation{%
  \institution{Software College, Northeastern University}
  \city{Shenyang}
  \country{China}
}
\email{zhaojz@swc.neu.edu.cn}

\author{Xingwei Wang}
\authornotemark[1]
\affiliation{%
  \institution{School of Computer Science and Engineering, Northeastern University}
  \city{Shenyang}
  \country{China}
}
\email{wangxw@mail.neu.edu.cn}

\renewcommand{\shortauthors}{Yizhou et al.}

\begin{abstract}
Sequential recommendation aims to provide users with personalized suggestions based on their historical interactions. When training sequential models, padding is a widely adopted technique for two main reasons: 1) The vast majority of models can only handle fixed-length sequences; 2) Batching-based training needs to ensure that the sequences in each batch have the same length. The special value \emph{0} is usually used as the padding content, which does not contain the actual information and is ignored in the model calculations. This common-sense padding strategy leads us to a problem that has never been explored before: \emph{Can we fully utilize this idle input space by padding other content to further improve model performance and training efficiency?} 

In this work, we propose a simple yet effective padding method called \textbf{Rep}eated \textbf{Pad}ding\textbf{+} (\textbf{RepPad+}). Specifically, we use the original interaction sequences as the padding content and fill it to the padding positions during model training. This operation can be performed a finite number of times or repeated until the input sequences' length reaches the maximum limit. For those sequences that can not pad full original data, we draw inspiration from the Sliding Windows strategy and intercept consecutive subsequences to fill in the idle space. Our RepPad+ can be viewed as a sequence-level data augmentation strategy. Unlike most existing works, our method contains no trainable parameters or hyperparameters and is a plug-and-play data augmentation operation. Extensive experiments on various categories of sequential models and seven real-world datasets demonstrate the effectiveness and efficiency of our approach. The average recommendation performance improvement is up to 84.11\% on GRU4Rec and 35.34\% on SASRec. We also provide in-depth analysis and explanation of what makes RepPad+ effective from multiple perspectives.
\end{abstract}

\ccsdesc[500]{Information systems~Recommender systems}

\keywords{Data Augmentation; Sequential Recommendation; Padding}


\maketitle

\section{Introduction}\label{sec:introduction}
As an essential branch of recommender systems, sequential recommendation (SR) has received much attention due to its well-consistency with real-world recommendation situations. It focuses on characterizing users' dynamic preferences in their historical sequences to predict the next user-item interaction(s). In the last few years, various sequential models have been proposed. Pioneering works employ Markov Chains \cite{rendle2010factorizing, shani2005mdp} and session-based KNN \cite{he2016fusing, hu2020modeling} to model sequential data. Later, CNN \cite{yuan2019simple, tang2018personalized} and RNN-based \cite{hidasi2015session, liu2016context, qu2022cmnrec} models are proposed to capture the dependencies among items within a sequence or across different sequences. More recently, the capability of Transformers \cite{vaswani2017attention, devlin2018bert} in encoding sequences has been adopted for applications in SR. Transformer-based SR models \cite{kang2018self, sun2019bert4rec, li2020time, dang2023uniform, dang2023ticoserec} encode sequences by learning item importance and behavior relevance via self-attention mechanisms.

\begin{figure}[!t]
	\centering
	\includegraphics[scale=0.48]{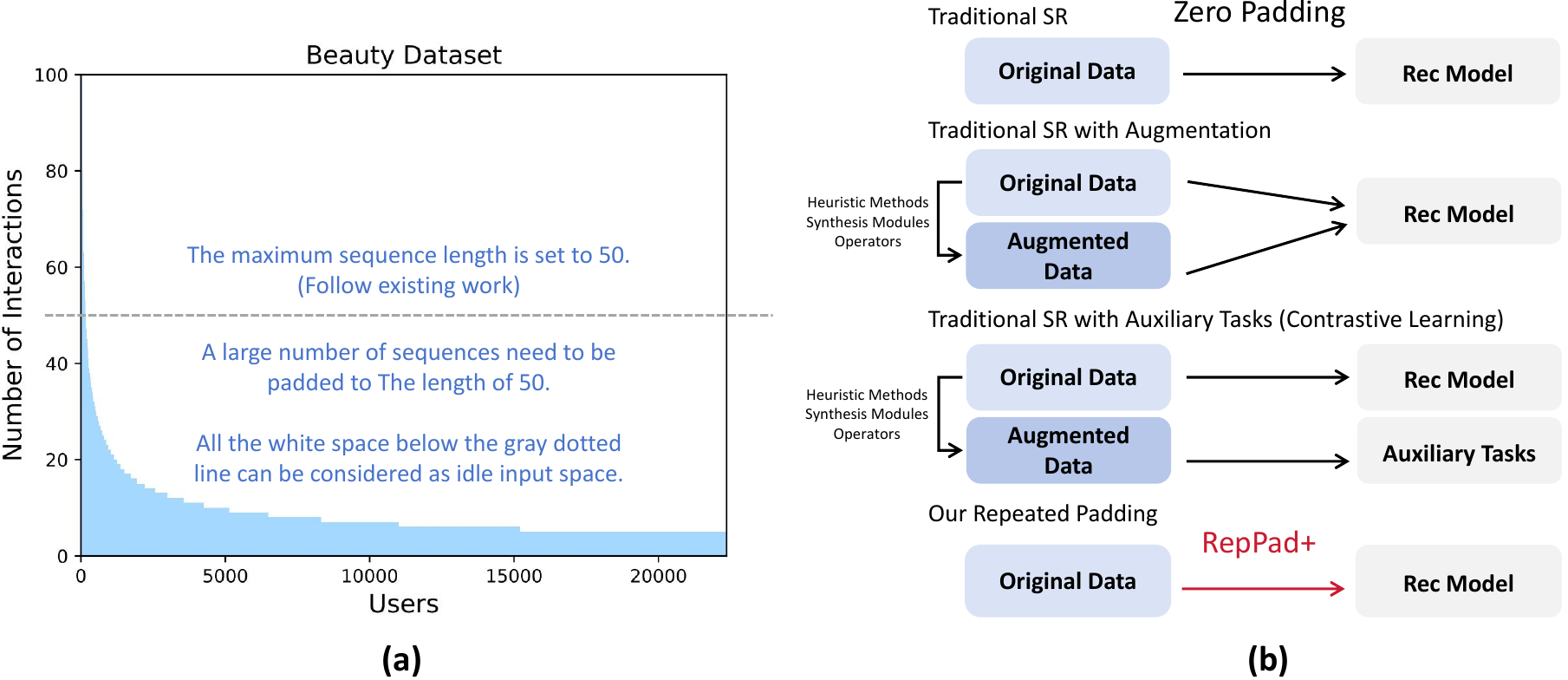}
	\caption{An illustration of (a) the idle input space in the sequential recommendation and (b) a comparison between traditional SR with padding and our repeated padding.}
	\label{fig:example}
\end{figure}

Although numerous models with different architectures have been proposed, padding has never been explored as an indispensable technique for training sequential models. The vast majority of sequence models can only support inputting fixed-length sequences to define the model architecture and process model calculations. In addition, batch processing techniques require all sequences in each batch to be the same length during training. Therefore, the maximum sequence length $N$ is usually used as a specified hyper-parameter before the model training. During training, interceptions are performed for long sequences with lengths larger than $N$, and padding is performed for short sequences with lengths smaller than $N$. In real-world online platforms, only a few users have long interaction sequences, while most users have very short ones. This phenomenon is also known as the long-tail effect. When padding a sequence, the special value \emph{0} is usually treated as content. It does not contain meaningful information and is not involved in model calculations. As illustrated in Figure \ref{fig:example} (a), this universally observed convention and custom leads us to the question: \emph{Can we fully utilize this idle input space by padding other content to improve model performance and training efficiency further?}

Motivated by the question above, we propose a simple yet effective padding strategy called \textbf{Rep}eated \textbf{Pad}ding\textbf{+} (\textbf{RepPad+} for short), which is based on the previous work RepPad proposed by Dang et al. \cite{dang2024repeated}. Unlike the traditional padding strategy using special value \emph{0}, we iteratively pad the original whole interaction sequence. This operation can be performed a limited number of times or repeated until the sequence length reaches a preset maximum limit. However, some sequences of moderate lengths can make it impossible to fill the remaining space with the complete original sequence. For example, when the original sequence length is 30 and $N=50$, the remaining space of length 20 cannot pad the original sequence. Inspired by the Sliding Windows strategy \cite{tang2018personalized}, we randomly intercept consecutive subsequences from the original sequence to fill these idle spaces. Besides, to avoid the situation where the head of the sequence is used to predict the tail, we add a delimiter between each padding. Our core idea is to utilize these idle input spaces that are not involved in model training but occupy positions to improve model performance and training efficiency.

From another perspective, RepPad+ is a sequence-level data augmentation method. The training of recommendation models relies heavily on large-scale labeled data. Data sparsity has been one of the major problems faced by recommender systems \cite{jing2023contrastive}. Earlier works use random data augmentation methods to increase the size of the training dataset directly, such as Dropout \cite{tan2016improved} and Slide Windows \cite{tang2018personalized}. Besides, some researchers utilize counterfactual thinking \cite{wang2021counterfactual}, bidirectional Transformer \cite{jiang2021sequential, liu2021augmenting}, or diffusion model \cite{liu2023diffusion} to augment sequential data. Inspired by recent developments of Contrastive Learning (CL) \cite{jing2023contrastive, yu2023self, zhao2023cross_1, zhao2023cross_2}, many effective data augmentation operators have been proposed \cite{xie2022contrastive, liu2021contrastive, dang2023uniform, qin2023intent, tian2023periodicity}. These models typically leverage shared encoders to encode original and augmented sequences. The representations of the original sequences are used for the main recommendation task, and the representations of the augmented sequence are used as the input for contrastive learning. It optimizes encoders by maximizing the agreements between ‘positive’ pairs, which are two augmentations from one sequence \cite{liu2021contrastive}. As illustrated in Figure \ref{fig:example} (b), these methods also use the special value \emph{0} as the padding content when dealing with sequences whose length is less than a specified value. Besides, adding additional sequence data to the original dataset makes the training process more time-consuming. Our RepPad+ utilizes the idle padding space to perform repetitive padding before the sequences are fed into the model. There is no additional increase in the number of sequences that need to be involved in the training. 

Different from elaborate data augmentation operators or the methods that require training, our method requires no training process and introduces no additional hyper-parameters or trainable parameters. It is independent of the backbone network and is a data augmentation plugin. We implement RepPad+ on various representative models based on different architectures. Extensive experimental results demonstrate that our approach can significantly improve the performance of sequential recommendation models based on various types of architectures. Regarding training efficiency and recommendation accuracy, RepPad+ achieves satisfactory improvements over existing heuristic and training-required data augmentation methods. Furthermore, we carefully analyze the different variants of RepPad+. To explore what makes RepPad+ effective, we provide an in-depth analysis from loss convergence, gradient stability, different padding strategies, and future information leakage perspectives.

Overall, our work makes the following contributions:
\begin{itemize}
  \item We leverage the idle input space of sequential models and propose a simple yet effective padding paradigm, RepPad+. Compared to RepPad, RepPad+ further considers and augments medium-length sequences.

  \item We compared RepPad+ in detail with existing methods. RepPad+ does not contain any parameters, requires no training process, does not increase the data amount in training, and is a plug-and-play augmentation method.

  \item We conduct comprehensive experiments on multiple real-world datasets, various architectures of sequential models, and augmentation methods. The results demonstrate the effectiveness and efficiency of our RepPad+. 
  
  \item We provide analysis and insights into why RepPad+ is effective from multiple perspectives. To our knowledge, this is the first work to explore the padding strategy in the recommendation system field.
\end{itemize}

\noindent \textbf{Outline.} The rest of this paper is organized as follows. Section 2 provides a brief overview of sequential recommendation models, padding methods in deep learning, and data augmentation for sequential recommendation in the literature. The previous work proposed by Dang et al. \cite{dang2024repeated}, and our proposed RepPad+ are introduced in Section 3. After that, the experimental results are presented and analyzed in Section 4, followed by the multi-perspective analysis of why our method works in Section 5. Finally, Section 6 concludes our work and outlines future research.

\section{Related Works}

\subsection{Sequential Recommendation}
Sequential recommendation (SR) is an emerging topic in the field of recommender systems \cite{qu2022cmnrec}. An early solution is to treat the item sequence as a Markov Chain \cite{rendle2010factorizing, shani2005mdp, he2016fusing}, where the next item to predict is closely related to the latest few interactions. The limitation lies in the inability to learn the dependency in a relatively large time step. Hence, a better solution based on recurrent neural networks (RNNs) \cite{schuster1997bidirectional} emerges since it can capture long- and short-term preferences. A pioneering work GRU4Rec \cite{hidasi2015session} introduced RNN into the field of recommender systems. It trained a Gated Recurrent Unit (GRU) architecture to model the evolution of user interests. Following this idea, CmnRec \cite{qu2022cmnrec} proposed to divide proximal information units into chunks, whereby the number of memory operations can be significantly reduced. However, RNN cannot be well paralleled and thus will reach its performance bottleneck when dealing with a large volume of sequential data. Another line of research is based on convolutional neural networks (CNNs). The representative work, Caser \cite{tang2018personalized}, embeds a sequence of recent items in the time and latent spaces, then learns sequential patterns using convolutional filters. NextItNet \cite{yuan2019simple} combines masked filters with 1D dilated convolutions to model the dependencies in interaction sequences. More recently, transformer-based approaches have attracted much attention because of their convenience for parallelization and carefully designed self-attention architecture \cite{vaswani2017attention, devlin2018bert}. SASRec \cite{kang2018self} employed unidirectional Transformers to fulfill the next-item prediction task in the sequential recommendation. Some works further optimized the Transformer's architecture to make it more suitable for recommendation scenarios, such as reducing computational complexity \cite{fan2021lighter} and stochastic self-attention \cite{fan2022sequential}. Furthermore, FMLP-Rec \cite{zhou2022filter} proposes an all-MLP architecture with learnable filters to enhance recommendation performance. Without exception, all of these methods use a zero-padding strategy when training the model. We argue that this strategy yields a huge amount of idle input space, which can be further utilized to improve model performance and training efficiency.

\subsection{Padding in Deep Learning}
In deep learning, "padding" refers to the process of adding extra pixels or values around the edges of input data, such as an image or a feature map. It is often used to handle boundary conditions and maintain the spatial dimensions or ensure consistent processing. Padding helps to preserve the integrity and context of the data during convolutional operations and other computations, and it can have a significant impact on the performance and output of deep learning models. For example, zero-padding is a common type where zeros are added. Currently, only a few works have explored the impact of padding on the deep learning models \cite{islam2021position}. An early work \cite{dwarampudi2019effects} analyzed the effects of different padding strategies on Long Short-Term Memory Networks and Convolutional Neural Networks. Further, some researchers demonstrated how the padding mechanism can induce spatial bias in CNNs in the form of skewed kernels and feature-map artifacts \cite{alsallakh2020mind}. In addition, some works proposed different padding strategies \cite{ning2023learning}, such as leverage partial convolution-based padding \cite{liu2018partial} to improve model performance and training efficiency, distribution padding \cite{nguyen2019distribution} to maintain the spatial distribution of the input border data, symmetric padding \cite{wu2019convolution} to achieve symmetry within a single convolution layer. The above works focused on the field of computer vision. In addition, the impact of padding in the field of natural language processing has also attracted attention \cite{gimenez2020semantic, nam2019padding, cheng2020empirical}, such as handwriting recognition \cite{nam2019padding} and text classification \cite{cheng2020empirical}. To our knowledge, we are the first work to explore padding strategy in the sequential recommendation and propose a simple yet effective method.

\subsection{Data Augmentation for Sequential Recommendation}
Data augmentation has been established as a standard technique to train more robust models in computer vision \cite{yang2022image, shorten2019survey} and natural language processing \cite{feng2021survey, li2022data}. In recommender systems, it has also been widely used and proved to be an effective method. Slide Windows is an early work \cite{tang2018personalized}, which modified the original click sequence by splitting it into many sub-sequences or randomly discarding a certain number of items. Similar to this simple operation, the noise/redundancy injection \cite{song2022data} and subset selection \cite{tan2016improved} have also been proven to be effective enhancement methods. Later, Conditional Generative Adversarial Nets were used to obtain reliable user interaction data from rich side information and further form augmented ground truth training datasets \cite{wang2019enhancing}. Inspired by counterfactual thinking, CASR \cite{wang2021counterfactual} revised the sequence of user behaviors by substituting some previously purchased items with other unknown items. Some researchers proposed to leverage bi-directional transformers \cite{liu2021augmenting, jiang2021sequential}. It employed a reversely pre-trained transformer to generate pseudo-prior items for short sequences. Then, fine-tune the pre-trained transformer to predict the next item. Besides, L2Aug \cite{wang2022learning} categorizes users into core and casual users, using historical data from the former to augment the interaction sequence of the latter. DiffuASR \cite{liu2023diffusion} leveraged the diffusion model for sequence generation. Two guide strategies are designed to make the generated items correspond to the raw data. With the rise of large language models (LLM), some researchers explored utilizing them as Data Augmenters to improve recommendations for cold-start items \cite{wang2024large}. 

In recent years, self-supervised learning techniques (e.g., contrastive learning) \cite{he2020momentum, chen2020simple} have shown great potential for alleviating the data sparsity problem. Many methods used it as an auxiliary task to further improve sequential model performance \cite{qin2023intent, xie2022contrastive}. For example, CL4SRec \cite{xie2022contrastive} came up with three data augmentation operators (Crop, Reorder, Mask) for pre-training recommendation tasks equipped with a contrastive learning framework. Based on CL4SRec, CoSeRec \cite{liu2021contrastive} integrated item similarity information for data augmentation (Insert, Substitute) with a contrastive learning objective, aiming to maximize the agreement of augmented sequences. To further improve the quality of the augmented sequence, some works performed augmentation concerning time and item category information \cite{dang2023ticoserec, tian2023periodicity}. In addition to sequence-level augmentation, augments and contrasts at the attribute level \cite{zhou2020s3}, embedding level \cite{qiu2022contrastive}, and model level \cite{hao2023learnable} has also been explored. Unlike existing methods, we are the first to explore the use of idle padding space for data augmentation. In terms of simplicity and applicability, our method does not contain any parameters or training processes and can be plugged into most existing sequential models.

\section{Methodology}
In this section, we formally present the problem of sequential recommendation (Section \ref{sec:Problem Formulation}). After that, we introduce the current widely used padding strategy (Section \ref{sec:Padding in Sequential Recommendation}) and briefly describe the sequence data augmentation (Section \ref{sec:Sequence Data Augmentation}). Next, we will present the previous work RepPad, which is proposed by Dang et al. \cite{dang2024repeated} (Section \ref{Repeated Padding (RepPad)}). Based on RepPad, we improve it by considering the medium-length sequences and further propose RepPad+ (Section \ref{RepPad+: Augmenting the Medium-length Sequences}). Finally, we provide a discussion between existing and our methods (Section \ref{Discussions}).

\subsection{Problem Formulation}\label{sec:Problem Formulation}
Suppose we have user and item sets denoted by symbols $\mathcal{U}$ and $\mathcal{V}$, respectively. Each user $u \in \mathcal{U}$ is associated with a sequence of interacted items in chronological order $s_u=[v_1, \ldots, v_j, \ldots, v_{\left|s_u\right|}]$, where $v_j \in \mathcal{V}$ indicate the item that user $u$ has interacted with at time step $j$ and $\left|s_u\right|$ is the sequence length. The sequential recommendation task can be formulated as follows. Given the sequences items $s_u$, how to accurately predict the most possible item $v^{*}$ that user $u$ will interact with at time step $\left|s_u\right|+1$, given by:
\begin{equation}
    \underset{v^{*} \in \mathcal{V}}{\arg \max} \;\; P\left(v_{\left|s_u\right|+1}=v^{*} \mid s_u \right).
\end{equation}
This equation can be interpreted as calculating the probability of all candidate items and selecting the highest one for the final recommendation.

\subsection{Padding in Sequential Recommendation}\label{sec:Padding in Sequential Recommendation}
\begin{figure}[!t]
	\centering
	\includegraphics[scale=0.38]{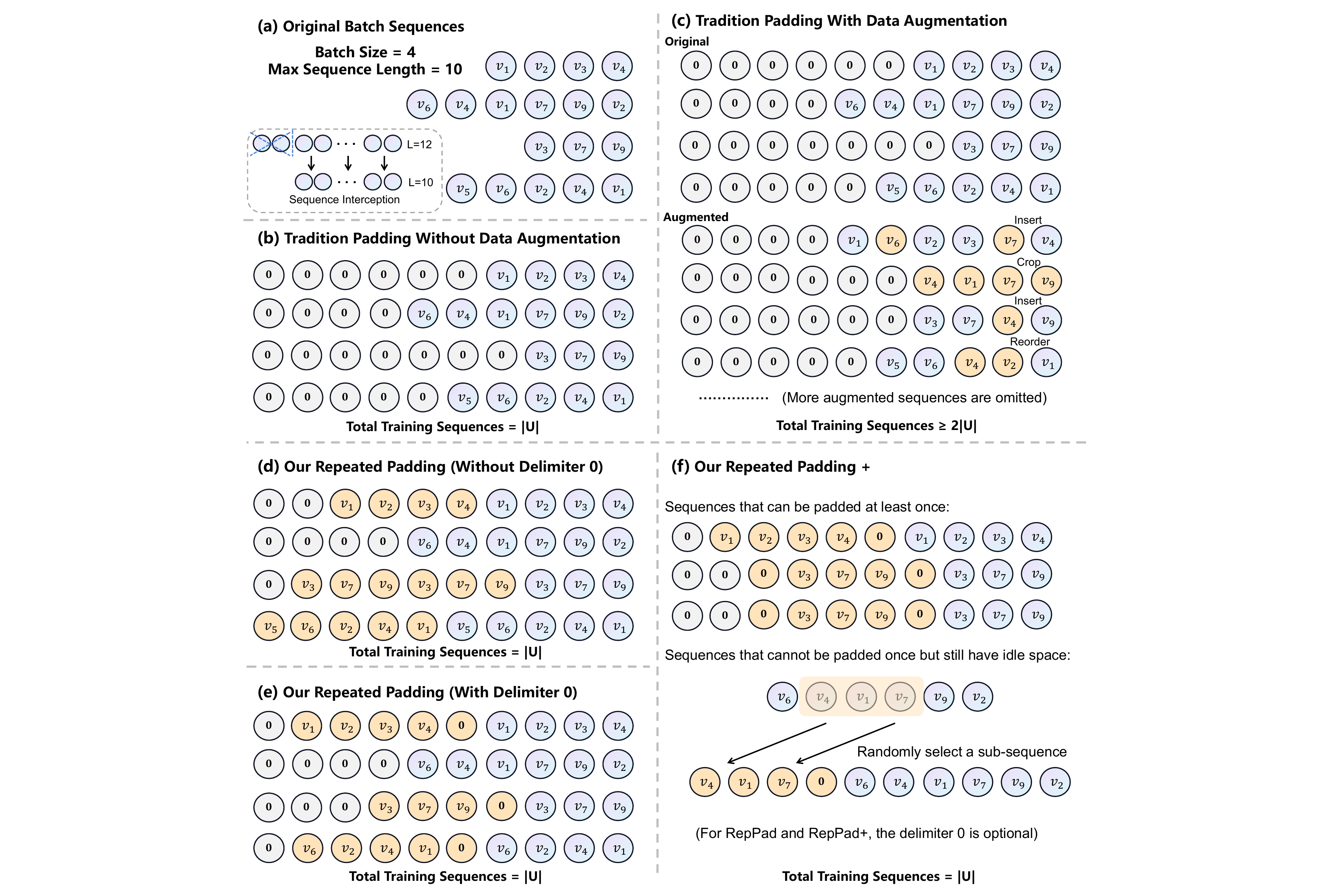}
	\caption{An illustration of (a) the original batch sequences, (b) the traditional padding, (c) traditional padding with data augmentation, (d,e) the previous work, RepPad, proposed by Dang et al. \cite{dang2024repeated}, and (f) our proposed RepPad+. We set the batch size = 4 and maximum sequence length = 10 for demonstration. The $\left|\mathcal{U}\right|$ is the number of users in the dataset.}
	\label{fig:frame}
\end{figure}

Padding is a common technique when training sequential models. The Batch process requires that the sequences in each batch have the same length. In addition, common CNN- or Transformer-based models can only handle fixed-length sequences and require model initialization and operations based on fixed sequence lengths. So, before training a recommendation model, we usually specify a maximum sequence length $N$. For sequences with an original length of more than $N$, the most recent $N$ items are usually chosen because they represent the user's recent preferences. As shown in Figure \ref{fig:frame} (a) and (b), original sequences of length less than $N$ are padded with the special value 0 to reach the length $N$. The special value 0 contains no actual information and does not participate in model calculations. The above process can be formulated as follows:
\begin{equation}
s_u^f = SeqPrepare\left(s_u\right) = \begin{cases}ZeroPad\left(s_u\right), & \left|s_u\right| < N \\ s_u, & \left|s_u\right|=N \\ Intercept\left(s_u\right), & \left|s_u\right|>N\end{cases}
\label{eq:pad}
\end{equation}
where $ZeroPad(\cdot)$ indicates padding the \emph{0} to the left side of the sequence until its length is $N$, and $Intercept(\cdot)$ indicates intercepting the nearest $N$ interactions. The $s_u^f$ is the final input to the model where $\small \left|s_u^f\right|= N$. Typically, the total number of sequences involved in model training is $\left|\mathcal{U}\right|$, i.e., one sequence for each user.

\subsection{Sequence Data Augmentation}\label{sec:Sequence Data Augmentation}
To alleviate the data sparsity problem, data augmentation is often used to augment the original data to increase the number of samples involved in model training. Given an original sequence $s_u$, a data augmentation operation makes minor but appropriate changes to $s_u$ to generate a new sequence $s_u^d$. Both $s_u$ and $s_u^d$ are used for model training or self-supervised learning. For example, as shown in Figure \ref{fig:frame} (c), for an original sequence $s_u=[v_5, v_2, v_6, v_4, v_1, v_7, v_9, v_2]$, we intercept a portion of the consecutive sequence to generate a new sequence data $s_u^d=[v_6, v_4, v_1, v_7, v_9]$. For the original sequence $s_u=[v_3, v_7, v_9]$, we inert an item $v_4$ between $v_7$ and $v_9$ to generate sequence data $s_u^d=[v_3, v_7, v_4, v_9]$. 

As previously described, this enhancement process can be accomplished by data augmentation operators, heuristics methods, and training-required synthetic modules. The generated data can be used directly to train recommendation models or for auxiliary tasks. In each epoch, each sequence will be augmented at least once, so the total number of sequences involved in model training will be at least 2$\left|\mathcal{U}\right|$. It is important to note that the original and augmented sequences also need to apply function Eq. (\ref{eq:pad}) before being fed into the model.

\subsection{The Previous Work: Repeated Padding (RepPad)}\label{Repeated Padding (RepPad)}
As illustrated in Figure \ref{fig:frame} (b) and (c), a large amount of idle space is filled with 0 regardless of whether data augmentation is performed. Also, due to the long-tail effect, the interaction sequences of the vast majority of users are usually short. Therefore, the idea of RepPad is to fully utilize these idle spaces further to improve the efficiency of model training and recommendation performance. 

In order to achieve this objective, previous work by Dang et al. \cite{dang2024repeated} proposes a simple but effective padding strategy called Repeated Padding (RepPad). Specifically, they utilize the original sequence as padding content instead of the special value \emph{0}. As shown in Figure \ref{fig:frame} (d), give the original sequence $s_u=[v_1, v_2, v_3, v_4]$, and they pad the original sequence once on the left and get the padded sequence $s_u^p=[v_1, v_2, v_3, v_4, v_1, v_2, v_3, v_4]$. This operation can be formulated as:
\begin{equation}
s_u^p = RepPad\left(m, s_u\right) = [\underbrace{s_u \;|\; s_u \;|\; \cdot\cdot\cdot \;|\; s_u}_{\text{$(m+1)~~s_u$}} ],
\label{eq:reppad}
\end{equation}
where $m$ is padding times and $|$ indicates sequence \en{concatenation}. This operation can be performed only once or repeated several times until there is not enough space left to hold the original sequence. In other words, $m$ can be pre-specified or calculated based on the original sequence length $s_u$ and the maximum sequence length $N$. For example, when the maximum sequence length is 10, the remaining length of 2 is insufficient to accommodate the original sequence of length 4, so it still needs to apply function Eq. (\ref{eq:pad}). The final input sequence $s_u^f=[0, 0, v_1, v_2, v_3, v_4, v_1, v_2, v_3, v_4]$. In addition, given the original sequence $s_u=[v_3, v_7, v_9]$, the padding operation can be performed up to 3 times, and the final sequence $s_u^f=[0, 0, v_3, v_7, v_9, v_3, v_7, v_9, v_3, v_7, v_9]$.

However, there is a problem with the above padding strategy: there may be cases where the end of the original sequence is used to predict the head. Using $[0, 0, \underline{v_1, v_2, v_3, v_4}, \underline{v_1}, v_2, v_3, v_4]$ as training data, there will be cases where $v_1, v_2, v_3, v_4$ is used to predict $v_1$. To solve this problem, they add the special value \emph{0} between each repeated padding to prevent this from happening. This operation can be formulated as:
\begin{equation}
s_u^p = RepPad\left(m, s_u\right) = [\underbrace{s_u \;|\; 0 \;|\; s_u \;|\; \cdots \;|\; 0 \;|\; s_u}_{\text{$(m+1)~~s_u$}}].
\label{eq:reppad_with0}
\end{equation}
As shown in Figure \ref{fig:frame} (e), given original sequence $s_u=[v_1, v_2, v_3, v_4]$, the final sequence after Eq. (\ref{eq:reppad_with0}) and (\ref{eq:pad}) is $[0, v_1, v_2, v_3, v_4, 0, v_1, v_2, v_3, v_4]$.  The possible values of $m$ and their interpretation are given in Table \ref{tab:valueofm}.

\begin{table}[!t]
\centering
 \caption{The value of $m$ and its interpretation.}
 \renewcommand\arraystretch{1.0}
\scalebox{1.0}{
 \begin{tabular}{c|c}
 \toprule \textbf{The Value of $m$} & \textbf{Interpretation} \\
 \midrule 
 $0$ & No repeated padding and use original sequence. \\
 $1, 2, 3, \cdots$ & Repeated padding a fixed number of times. \\
 $max$ & Padding until the remaining space is insufficient.  \\
 $random(0, max)$ & Randomly select between 0 and maximum value.\\
 $random(1, max)$ & Randomly select between 1 and maximum value.\\
 \bottomrule
 \end{tabular}}
 \label{tab:valueofm}
\end{table}

\begin{algorithm}[!t]
  \caption{RepPad Pseudocode ($m = random(1, max)$)}
  \label{alg:code}
  \definecolor{codeblue}{rgb}{0.25,0.5,0.5}
  \definecolor{codekw}{rgb}{0.85, 0.18, 0.50}
  \lstset{
    backgroundcolor=\color{white},
    basicstyle=\fontsize{7.5pt}{7.5pt}\ttfamily\selectfont,
    columns=fullflexible,
    breaklines=true,
    captionpos=b,
    commentstyle=\fontsize{7.5pt}{7.5pt}\color{codeblue},
    keywordstyle=\fontsize{7.5pt}{7.5pt}\color{codekw},
}
  \begin{lstlisting}[language=python, mathescape]
  # orig_seq: An Original User Sequence s_u.
  # max_len: Maximum Sequence Length N.
  # The leave-one-out strategy is adopted. The last item (origseq[:-1]) in each sequence is the test set, and the penultimate item (orig_seq[:-2]) is the validation set.

  def reppad(orig_seq, max_len):
    final_input = orig_seq[:-3]  # training items
    final_target = orig_seq[1:-2]  # target items
    
    if int(max_len / len(final_input)) <= 1:
      # Insufficient space, no padding operation.
      return final_input, final_target
    
    else:
      # Calculate the maximum padding count.
      max_pad_num = int(max_len / len(final_input))

      # Randomly select a padding count.
      pad_num = random.randint(1, max_pad_num)
      
      # Perform the Repeated Padding operation.
      final_input = (final_input + [0]) * pad_num + final_input
      final_target = (final_target + [0]) * pad_num + final_target
      return final_input, final_target

  # Then apply function Eq.(2) before being fed into the model.
  \end{lstlisting}
\end{algorithm}

The Python procedure of RepPad is presented in Algorithm \ref{alg:code}. So far, the basic idea of RepPad has been introduced. It is worth emphasizing that the repeated padding sequences still need to apply function Eq. (\ref{eq:pad}) operation before they can be fed into the model:
\begin{equation}
s_u^f = SeqPrepare\left(s_u^p\right).
\label{eq:final}
\end{equation}

\subsection{RepPad+: Augmenting the Medium-length Sequences}\label{RepPad+: Augmenting the Medium-length Sequences}
In Figure \ref{fig:example} (d) and (e), we can observe that sequence $[0, 0, 0, 0, v_6, v_4, v_1, v_7, v_9, v_2]$ cannot be augmented because of insufficient remaining space. However, for this sequence, there is still idle input space of length 4. Naturally, we can consider padding this space using consecutive sub-sequences of the original sequence. However, if we follow the general interception paradigm, i.e., only the most recent interactions are intercepted, it may cause earlier interactions to be ignored in training, resulting in inaccurate representations learning. Inspired by Sliding Windows \cite{tang2018personalized}, We propose to randomly truncate subsequences from the original sequence for padding. Specifically, we first calculate the length of the remaining space based on the maximum sequence length $N$ and the length of the original sequence $L$. The length of the remaining space is equal to $N-L-1$ if delimiter 0 is used, otherwise $N-L$. Let's take the case of using the delimiter 0 as an example. Next, we randomly truncate the sub-sequence from the original sequence based on this length:
\begin{equation}
s_u^{sub} = subseq\left(s_u, N-L-1\right).
\label{eq:sub}
\end{equation}
Then, we use $s_u^{sub}$ as the padding content:
\begin{equation}
s_u^p = \underbrace{[s_u^{sub} \;|\; 0 \;|\; s_u]}_{Length=N.}.
\label{eq:RepPad+}
\end{equation}
It should be noted that the above operation is only applicable to sequences where the remaining space is insufficient for padding the whole original sequence, and the delimiter 0 is optimal. As illustrated in Figure \ref{fig:frame} (f), this padding operation is only applied to sequence $[0, 0, 0, 0, v_6, v_4, v_1, v_7, v_9, v_2]$. Applying Eq. \ref{eq:sub} we get sub-sequence $[v_4, v_1, v_7]$. Through Eq. \ref{eq:RepPad+}, the final padded sequence is $[v_4, v_1, v_7, 0, v_6, v_4, v_1, v_7, v_9, v_2]$ through Eq. \ref{eq:RepPad+}.

Based on the RepPad in Section \ref{Repeated Padding (RepPad)}, we add this operation and refer to this version as RepPad+. As mentioned in Section \ref{Repeated Padding (RepPad)}, in practice, we strongly recommend users apply function Eq. (\ref{eq:pad}) after using our method, which can be seen as a final check on sequence length. Due to the influence of different training data and code implementations, the length of the augmented sequence may not be $N$, and using Eq. (\ref{eq:pad}) can ensure that the length of the sequence meets the conditions to be fed into the model. We present the Python procedure of RepPad+ in Algorithm \ref{alg:code+}. Note that for RepPad+, since medium-length sequences are not suitable for full repeated padding, the number of $m$ has no effect on their augmentation. Only whether or not the delimiter \emph{0} is added (i.e., whether there is a delimiter \emph{0} between the padding sub-sequence and the original sequence) will have an effect on the augmentation of medium-length sequences.

\begin{algorithm}[!t]
  \caption{RepPad+ Pseudocode ($m = random(1, max)$)}
  \label{alg:code+}
  \definecolor{codeblue}{rgb}{0.25,0.5,0.5}
  \definecolor{codekw}{rgb}{0.85, 0.18, 0.50}
  \lstset{
    backgroundcolor=\color{white},
    basicstyle=\fontsize{7.5pt}{7.5pt}\ttfamily\selectfont,
    columns=fullflexible,
    breaklines=true,
    captionpos=b,
    commentstyle=\fontsize{7.5pt}{7.5pt}\color{codeblue},
    keywordstyle=\fontsize{7.5pt}{7.5pt}\color{codekw},
    }
  \begin{lstlisting}[language=python, mathescape]
  # orig_seq: An Original User Sequence s_u.
  # max_len: Maximum Sequence Length N.
  # The leave-one-out strategy is adopted. The last item (origseq[:-1]) in each sequence is the test set, and the penultimate item (orig_seq[:-2]) is the validation set.

  def reppad+(orig_seq, max_len):
    final_input = orig_seq[:-3]  # training items
    final_target = orig_seq[1:-2]  # target items

    if len(orig_seq[:-3]) > max_len - 2:
      # No space for padding operation. One position for delimiter 0, One position for an item.
      return final_input, final_target
    
    elif int(max_len / len(final_input)) <= 1:
      # Not enough space left for padding whole original sequence. Padding sub-sequence.
      sub_sequence_len = max_len - 1 - len(final_input)
      start_index = random.randint(0, len(final_input) - sub_sequence_len)
      final_input = final_input[start_index:start_index + sub_sequence_len] + [0] + final_input
      final_target = final_target[start_index:start_index + sub_sequence_len] + [0] + final_target
      return final_input, final_target
    
    else:
      # Padding original sequence.
      # Calculate the maximum padding count.
      max_pad_num = int(max_len / len(final_input))

      # Randomly select a padding count.
      pad_num = random.randint(1, max_pad_num)
      
      # Perform the Repeated Padding operation.
      final_input = (final_input + [0]) * pad_num + final_input
      final_target = (final_target + [0]) * pad_num + final_target
      return final_input, final_target

  # Then apply function Eq.(2) before being fed into the model.
  \end{lstlisting}
\end{algorithm}

\subsection{Discussions}\label{Discussions}
\subsubsection{\textbf{Comparison with Existing Methods.}} We analyze the advantages of our approach over existing methods from the following perspectives: 
\begin{itemize}
    \item[] (\romannumeral 1) \textbf{\textit{No increase in training data size}}: From Figure \ref{fig:frame}, we can observe that our RepPad+ performs data augmentation for both short and medium sequences without increasing the amount of data for training. While previous approaches can be viewed as extending the length of the dataset vertically, our approach utilizes the idle space horizontally. For traditional data augmentation methods \cite{tan2016improved, tang2018personalized, wang2021counterfactual}, each sequence will be augmented at least once in each epoch, so the amount of data will be at least $2\left|\mathcal{U}\right|$. Some methods augment specific sequences (e.g., cold start user sequence) in the dataset \cite{liu2021augmenting, wang2024large, wang2022learning}, so the total number of sequences will be at least greater than $\mathcal{U}$. For sequential recommendation models that leverage both data augmentation and contrastive learning \cite{xie2022contrastive, liu2021contrastive, dang2023uniform}, each sequence is augmented at least twice, and with the addition of the original sequence, the number of model sequences involved in training is at least $3\left|\mathcal{U}\right|$. In addition, according to the optimal parameter settings of some methods \cite{liu2023diffusion, yin2024dataset}, the total number of sequences in a single epoch may reach $4\left|\mathcal{U}\right|$ or $5\left|\mathcal{U}\right|$.

    \item[] (\romannumeral 2) \textbf{\textit{No hyper-parameters and no increase in model parameters}}: These existing augmentation methods or augmentation operators (Slide Window, Dropout, Crop, Insert, Mask, Substitute, Reorder, and so on) typically contain at least one hyper-parameter for each that must be manually tuned based on the dataset \cite{tan2016improved, tang2018personalized, xie2022contrastive, liu2021contrastive}. Some training-required data generation modules introduce additional model parameters in addition to hyper-parameters \cite{wang2022learning, liu2023diffusion,yin2024dataset}. However, our approach does not involve any hyper-parameters that need to be tuned carefully, nor does it increase the number of parameters in the original model.

    \item[] (\romannumeral 3) \textbf{\textit{No training cost and model agnostic}}: Some researchers propose to use bidirectional Transformers \cite{liu2021augmenting, jiang2021sequential}, diffusion models \cite{liu2023diffusion} or encoder-decoder based augmenter \cite{yin2024dataset} for data augmentation. These methods require training of data augmentation modules in addition to training of backbone recommendation networks and have limitations on the type of backbone network. In contrast, our method does not require extra training processes and can be adapted to most sequential models. We present a summary comparison between PepPad+ and existing methods in Table \ref{tab:comparison}. It should be emphasized that our focus is on sequence-level data augmentation, which is also the most mainstream practice in data augmentation in the field of sequence recommendation. Some works explore augmentation at the embedding \cite{qiu2022contrastive} and model \cite{hao2023learnable} levels. These methods do not operate directly on the sequence data and are not involved in our discussion.
\end{itemize}

\begin{table}[!t]
	\centering
 	\caption{Comparison of sequence-level data augmentation between our method and several representative methods. The `No HP \& MP' represents no additional hyper-parameters or increases in model parameters. The `No Training' represents no training process for data generation modules or auxiliary tasks. The `TTS'  represents the total training sequences in each epoch.}
        \renewcommand\arraystretch{1.0}
	\scalebox{0.75}{
		\begin{tabular}{l|ccccc}
		\toprule \textbf{{Methods}} & \textbf{Augmentation Type} &\textbf{No HP \& MP} & \textbf{No Training} & \textbf{Model Agnostic} & \textbf{TTS} \\
			\midrule 
			{Only Augmentation \cite{tan2016improved, tang2018personalized}} & Random or Elaborate Operators & \ding{56} & \ding{52} & \ding{52} & $\geq$ $2\left|\mathcal{U}\right|$\\
			 {Augmentation with CL \cite{xie2022contrastive, liu2021contrastive, dang2023uniform, qin2023intent, tian2023periodicity}} & Operators or Synthesis Modules & \ding{56} & \ding{56} & \ding{52} & $\geq$ $3\left|\mathcal{U}\right|$\\
			 {CASR \cite{wang2021counterfactual}} & Counterfactual Thinking & \ding{56} & \ding{56} & \ding{52} & $\geq$ $2\left|\mathcal{U}\right|$\\
    {ASReP \cite{liu2021augmenting} and BiCAT \cite{jiang2021sequential}} & Bidirectional Transformers &  \ding{56} & \ding{56} & \ding{56} & $>$ $\left|\mathcal{U}\right|$\\
    {L2Aug \cite{wang2022learning}} & Synthesis Modules &  \ding{56} & \ding{56} & \ding{52} & $>$ $\left|\mathcal{U}\right|$\\
    {LLM Augmentation \cite{wang2024large}} & Large Language Models & \ding{56} & \ding{56} & \ding{52} & $>$ $\left|\mathcal{U}\right|$\\
    {DiffuASR \cite{liu2023diffusion}} & Diffusion Models & \ding{56} & \ding{56} & \ding{52} & $\geq$ $4\left|\mathcal{U}\right|$\\
    RepPad \cite{dang2024repeated} & Repeated Padding for short & \ding{52} & \ding{52} & \ding{52} & $=$ $\left|\mathcal{U}\right|$\\
    \textbf{RepPad+ (Ours)} & Repeated Padding for short and medium & \ding{52} & \ding{52} & \ding{52} & $=$ $\left|\mathcal{U}\right|$\\
			\bottomrule
	\end{tabular}}
	\label{tab:comparison}
\end{table}

\subsubsection{\textbf{Possible Limitations.}}\label{sec:Limitations} Our approach intends to maximize the use of idle input space. Intuitively, it is more effective on short-sequence datasets and may be ineffective on long-sequence datasets. In long-sequence datasets, most users have sequence lengths that close to or exceed the given maximum sequence length, and thus, there will be no space left for repeated padding. For RepPad and our proposed RepPad+, the latter will outperform the former on long sequence datasets. This is because RepPad+ performs augmentation on both short and medium-length sequences, whereas the RepPad performs only on short sequences. As shown in Figure \ref{fig:example} (a), only a very small fraction of users (0.72\% in the Beauty dataset) have sequences longer than the maximum sequence length $N$. We will verify our conjecture in the experimental section.

\section{Experiments}\label{sec:Experiments}

\subsection{Experiment Setup}
\subsubsection{Datasets} The experiments are conducted on seven widely-used public datasets:
\begin{itemize}[leftmargin=*]
  \item \textbf{Toys, Beauty, Sports, and Home:} Four datasets obtained from Amazon \cite{mcauley2015inferring} and correspond to the "Toys and Games", "All Beauty, "Sports and Outdoors", and "Home and Kitchen" categories, respectively.

  \item \textbf{Yelp\footnote{\url{https://www.yelp.com/dataset}}:} This is a popular dataset for business recommendation. As it is very large, we only use the transaction records after January 1st, 2019.

  \item \textbf{LastFM\footnote{\url{https://grouplens.org/datasets/hetrec-2011/}}:} It is a music artist recommendation dataset and contains user tagging behaviors for artists.
  
  \item \textbf{MovieLens-1M\footnote{\url{https://grouplens.org/datasets/movielens/}} \en{(ML-1M for short)}:} It comprises rating behaviors gathered from a movie review website.
\end{itemize}

The Amazon and Yelp datasets contain mainly short user sequences. The MovieLens-1M and LastFM datasets contain mainly long user sequences. Following \cite{liu2021contrastive, dang2023ticoserec, tian2023periodicity}, we reduce the data by extracting the 5-core. The maximum sequence length $N$ is set to 200 for the MovieLens-1M dataset, 100 for LastFM dataset, and 50 for other datasets. The detailed statistics are summarized in Table \ref{tab:datasets}.

\begin{table}[!t]
  \centering
 	\caption{The statistics of seven datasets. The `AL' denotes average length.}
  \scalebox{1.0}{
    \begin{tabular}{l|rrrrr}
    \toprule
    \textbf{Dataset} & \textbf{\# Users} & \textbf{\# Items} & \textbf{\# Interactions} & \textbf{\# AL} & \textbf{Sparsity} \\
    \midrule
    LastFm & 1,090 & 3,646 & 52,551 & 48.2 & 98.68\% \\
    Toys  & 19,412 & 11,924 & 167,597 & 8.6 & 99.93\% \\
    Beauty & 22,363 & 12,101 & 198,502 & 8.9 & 99.93\% \\
    Sports & 35,598 & 18,357 & 296,337 & 8.3 & 99.95\% \\
    Yelp  & 30,431 & 20,033 & 316,354 & 10.4 & 99.95\% \\
    Home  & 66,519 & 28,237 & 551,682 & 8.3 & 99.97\% \\
    ML-1M & 6,040 & 3,706 & 1,000,209 & 165.6 & 95.53\% \\
    \bottomrule
    \end{tabular}}%
  \label{tab:datasets}%
\end{table}%

\subsubsection{Baselines} To provide a comprehensive evaluation of RepPad+, the baselines consist of three main categories:

\noindent \textit{\textbf{First}} category is eight representative sequential recommendation models, which are used to evaluate whether RepPad+ is effective in improving the model's performance. 
\begin{itemize}[leftmargin=*]
  \item RNN-based models: \textbf{GRU4Rec} \cite{hidasi2015session} leverages gated recurrent units to model behavioral patterns. It argues that by modeling the whole session, more accurate recommendations can be provided. \textbf{NARM} \cite{li2017neural} incorporates attention mechanism into RNN architectures. It proposes to capture both the user's sequential behavior and main purpose in the current session.

  \item CNN-based models: \textbf{Caser} \cite{tang2018personalized} models recent actions as an "image" among time and latent dimensions. It learns sequential patterns using convolutional filters. \textbf{NextItNet} \cite{yuan2019simple} combines masked filters with 1D dilated convolutions to increase the receptive fields, which is important to model the long-range dependencies.

  \item MLP-based models: \textbf{gMLP} suggests a simpler alternative to the multi-head self-attention layers in Transformers. It adopts a cross-token MLP structure named Spatial Gating Unit to capture sequential information. \textbf{FMLP-Rec} \cite{zhou2022filter} finds that the logged user behavior data usually contains noisy interactions, and filtering algorithms from the digital signal processing area are useful to alleviate the influence of the noise in deep SR models. Inspired by this finding, an all-MLP model with learnable filters for SR tasks is proposed.

  \item Transformer-based models: \textbf{SASRec} \cite{kang2018self} is the representative work in a sequential recommendation that leverages the multi-head attention mechanism to the next item recommendation. It models the entire user sequence (without any recurrent or convolutional operations). \textbf{LightSANs} \cite{fan2021lighter} introduces the low-rank decomposed self-attention, which projects the user's historical items into a small constant number of latent interests and leverages item to-interest interaction to generate the context-aware representation. They propose overcoming quadratic complexity and over-parameterization. 
\end{itemize}

\noindent \textit{\textbf{Second}} category is the widely used stochastic data augmentation methods, which does not require training but include several hyper-parameters that need to be manually tuned. 
\begin{itemize}[leftmargin=*]
  \item \textbf{Random (Ran)} \cite{liu2023diffusion}: This method augments each sequence by randomly selecting items from the whole item set $\mathcal{V}$.

  \item \textbf{Slide Window (SW)} \cite{tang2018personalized}: It adopts slide windows to intercept multiple new subsequences from the original data.

  \item \textbf{Random-seq (Ran-S)} \cite{liu2023diffusion}: It selects items from the original sequence randomly as the augmentation items.

  \item \textbf{CMR} \cite{xie2022contrastive}: It proposes three augmentation operators with contrastive learning. In this category, we only use the three operators, including Crop, Mask, and Reorder.

  \item \textbf{CMRSI} \cite{liu2021contrastive}: Based on CMR, this work proposes two informative operators. We adopt five operators including Crop, Mask, Reorder, Substitute, and Insert.
\end{itemize}

\noindent \textit{\textbf{Third}} category is representative of training-required augmentation models, which usually include trainable parameters and hyper-parameters that need to be manually tuned. 
\begin{itemize}[leftmargin=*]
  \item \textbf{ASReP} \cite{liu2021augmenting}: It employs a reversely pre-trained transformer to generate pseudo-prior items for short interaction sequences. Then, fine-tune the pre-trained transformer to predict the next item.

  \item \textbf{DiffuASR} \cite{liu2023diffusion}: It adopts the diffusion model for discrete sequence generation tasks. Besides, two guide strategies are designed to make the generated items correspond to the raw data.

  \item \textbf{CL4SRec} \cite{xie2022contrastive}: This method leverages random data augmentation operators with contrastive learning to extract self-supervised signals from the original data. 
\end{itemize}
We do not include methods L2Aug \cite{wang2022learning}, CASR \cite{wang2021counterfactual}, BiCAT \cite{jiang2021sequential}, and LLM Augmentation \cite{wang2024large} since they do not provide open-source codes for reliable reproduction.

\subsubsection{Evaluation Metrics} Following \cite{liu2021contrastive, dang2023ticoserec, tian2023periodicity}, we adopt the leave-one-out strategy, wherein the last item of each user sequence serves as the test data, the items preceding it as validation data, and the remaining data as training data. We rank the prediction over the whole item set rather than negative sampling, otherwise leading to biased discoveries \cite{krichene2020sampled}. The evaluation metrics include Hit Ratio@\en{K} (denoted by H@\en{K}), and Normalized Discounted Cumulative Gain@\en{K} (N@\en{K}). We report results with \en{K} $\in \{5, 10, 20\}$. Generally, \emph{greater} values imply \emph{better} ranking accuracy.

\subsubsection{Implementation Details} For all baselines, we adopt the implementation provided by the authors. We set the embedding size to 64 and the batch size to 256 for all methods. To ensure fair comparisons, we carefully set and tune all other hyper-parameters of each method as reported and suggested in the original papers. We use the Adam \cite{kingma2014adam} optimizer with a learning rate of 0.001, $\beta_1=0.9$, $\beta_2=0.999$. For all methods, we adopt early stopping on the validation set if the performance does not improve for 20 epochs and report results on the test set. Our method does not contain any parameters that require additional tuning. For a fair comparison, the hyper-parameters of the original model will remain the same before and after adding RepPad and RepPad+.

\begin{table}[ht]
  \centering
  \caption{Performance improvement on five short-sequence datasets. The `w/ RP' and `w/ RP+' represent adding RepPad and Our RepPad+ to the corresponding model, respectively. The best performance of the three is bolded. The `Imp-O' and `Imp-R' represent the relative improvements of RepPad+ over the original model and the RepPad, respectively. All improvements are statistically significant, as determined by a paired t-test with $p \leq 0.05$.}
  \renewcommand\arraystretch{1.0}
  \setlength{\tabcolsep}{1mm}{
  \scalebox{0.59}{
    \begin{tabular}{cc|ccccc|ccccc|ccccc|ccccc}
    \toprule
    \multicolumn{1}{c|}{Dataset} & Metric & GRU4Rec & w/ RP & w/ RP+ & Imp-O & Imp-R & NARM & w/ RP & w/ RP+ & Imp-O & Imp-R & Caser & w/ RP & w/ RP+ & Imp-O & Imp-R & NextItNet & w/ RP & w/ RP+ & Imp-O & Imp-R \\
    \midrule 
    
    \multicolumn{1}{c|}{\multirow{6}[0]{*}{Toys}} & H@5 & 0.0134 & 0.0276 & \textbf{0.0319} & 138.06\% & 15.58\% & 0.0275 & 0.0286 & \textbf{0.0309} & 12.36\% & 8.04\% & 0.0132 & 0.0152 & \textbf{0.0173} & 31.06\% & 13.82\% & 0.0229 & 0.0273 & \textbf{0.0292} & 27.51\% & 6.96\% \\
    \multicolumn{1}{c|}{} & H@10 & 0.0256 & 0.0434 & \textbf{0.0482} & 88.28\% & 11.06\% & 0.0401 & 0.0415 & \textbf{0.0439} & 9.48\% & 5.78\% & 0.0231 & 0.0246 & \textbf{0.0269} & 16.45\% & 9.35\% & 0.0391 & 0.0452 & \textbf{0.0463} & 18.41\% & 2.43\% \\
    \multicolumn{1}{c|}{} & H@20 & 0.0432 & 0.0652 & \textbf{0.0716} & 65.74\% & 9.82\% & 0.0598 & 0.0589 & \textbf{0.0611} & 2.17\% & 3.74\% & 0.0364 & 0.0380 & \textbf{0.0403} & 10.71\% & 6.05\% & 0.0539 & 0.0613 & \textbf{0.0630} & 16.88\% & 2.77\% \\
    \multicolumn{1}{c|}{} & N@5 & 0.0083 & 0.0179 & \textbf{0.0217} & 161.45\% & 21.23\% & 0.0179 & 0.0188 & \textbf{0.0208} & 16.20\% & 10.64\% & 0.0079 & 0.0096 & \textbf{0.0109} & 37.97\% & 13.54\% & 0.0128 & 0.0153 & \textbf{0.0161} & 25.78\% & 5.23\% \\
    \multicolumn{1}{c|}{} & N@10 & 0.0121 & 0.0229 & \textbf{0.0269} & 122.31\% & 17.47\% & 0.0220 & 0.0233 & \textbf{0.0251} & 14.09\% & 7.73\% & 0.0111 & 0.0126 & \textbf{0.0145} & 30.63\% & 15.08\% & 0.0172 & 0.0201 & \textbf{0.0219} & 27.33\% & 8.96\% \\
    \multicolumn{1}{c|}{} & N@20 & 0.0166 & 0.0284 & \textbf{0.0328} & 97.59\% & 15.49\% & 0.0269 & 0.0276 & \textbf{0.0294} & 9.29\% & 6.52\% & 0.0149 & 0.0160 & \textbf{0.0177} & 18.79\% & 10.63\% & 0.0217 & 0.0251 & \textbf{0.0263} & 21.20\% & 4.78\% \\
    \midrule
    
    \multicolumn{1}{c|}{\multirow{6}[0]{*}{Beauty}} & H@5 & 0.0182 & 0.0286 & \textbf{0.0306} & 68.13\% & 6.99\% & 0.0213 & 0.0244 & \textbf{0.0283} & 32.86\% & 15.98\% & 0.0149 & 0.0199 & \textbf{0.0215} & 44.30\% & 8.04\% & 0.0223 & 0.0263 & \textbf{0.0269} & 20.63\% & 2.28\% \\
    \multicolumn{1}{c|}{} & H@10 & 0.0327 & 0.0468 & \textbf{0.0505} & 54.43\% & 7.91\% & 0.0367 & 0.0400 & \textbf{0.0447} & 21.80\% & 11.75\% & 0.0261 & 0.0330 & \textbf{0.0340} & 30.27\% & 3.03\% & 0.0393 & 0.0442 & \textbf{0.0456} & 16.03\% & 3.17\% \\
    \multicolumn{1}{c|}{} & H@20 & 0.0551 & 0.0732 & \textbf{0.0756} & 37.21\% & 3.28\% & 0.0617 & 0.0659 & \textbf{0.0682} & 10.53\% & 3.49\% & 0.0457 & 0.0511 & \textbf{0.0525} & 14.88\% & 2.74\% & 0.0633 & 0.0725 & \textbf{0.0743} & 17.38\% & 2.48\% \\
    \multicolumn{1}{c|}{} & N@5 & 0.0116 & 0.0175 & \textbf{0.0198} & 70.69\% & 13.14\% & 0.0131 & 0.0152 & \textbf{0.0187} & 42.75\% & 23.03\% & 0.0090 & 0.0127 & \textbf{0.0139} & 54.44\% & 9.45\% & 0.0135 & 0.0151 & \textbf{0.0157} & 16.30\% & 3.97\% \\
    \multicolumn{1}{c|}{} & N@10 & 0.0162 & 0.0234 & \textbf{0.0263} & 62.35\% & 12.39\% & 0.0180 & 0.0202 & \textbf{0.0240} & 33.33\% & 18.81\% & 0.0126 & 0.0169 & \textbf{0.0179} & 42.06\% & 5.92\% & 0.0191 & 0.0221 & \textbf{0.0235} & 23.04\% & 6.33\% \\
    \multicolumn{1}{c|}{} & N@20 & 0.0219 & 0.0300 & \textbf{0.0326} & 48.86\% & 8.67\% & 0.0243 & 0.0258 & \textbf{0.0294} & 20.99\% & 13.95\% & 0.0175 & 0.0214 & \textbf{0.0223} & 27.43\% & 4.21\% & 0.0251 & 0.0280 & \textbf{0.0289} & 15.14\% & 3.21\% \\
    \midrule
    
    \multicolumn{1}{c|}{\multirow{6}[0]{*}{Sports}} & H@5 & 0.0078 & 0.0114 & \textbf{0.0133} & 70.51\% & 16.67\% & 0.0126 & 0.0140 & \textbf{0.0157} & 24.60\% & 12.14\% & 0.0081 & 0.0092 & \textbf{0.0101} & 24.69\% & 9.78\% & 0.0104 & 0.0112 & \textbf{0.0123} & 18.27\% & 9.82\% \\
    \multicolumn{1}{c|}{} & H@10 & 0.0143 & 0.0204 & \textbf{0.0232} & 62.24\% & 13.73\% & 0.0218 & 0.0238 & \textbf{0.0249} & 14.22\% & 4.62\% & 0.0146 & 0.0160 & \textbf{0.0165} & 13.01\% & 3.13\% & 0.0195 & 0.0215 & \textbf{0.0234} & 20.00\% & 8.84\% \\
    \multicolumn{1}{c|}{} & H@20 & 0.0249 & 0.0338 & \textbf{0.0383} & 53.82\% & 13.31\% & 0.0361 & 0.0385 & \textbf{0.0398} & 10.25\% & 3.38\% & 0.0257 & 0.0265 & \textbf{0.0272} & 5.84\% & 2.64\% & 0.0338 & 0.0368 & \textbf{0.0386} & 14.20\% & 4.89\% \\
    \multicolumn{1}{c|}{} & N@5 & 0.0044 & 0.0073 & \textbf{0.0083} & 88.64\% & 13.70\% & 0.0076 & 0.0088 & \textbf{0.0103} & 35.53\% & 17.05\% & 0.0048 & 0.0056 & \textbf{0.0068} & 41.67\% & 21.43\% & 0.0064 & 0.0074 & \textbf{0.0079} & 23.44\% & 6.76\% \\
    \multicolumn{1}{c|}{} & N@10 & 0.0065 & 0.0100 & \textbf{0.0115} & 76.92\% & 15.00\% & 0.0106 & 0.0120 & \textbf{0.0132} & 24.53\% & 10.00\% & 0.0069 & 0.0078 & \textbf{0.0089} & 28.99\% & 14.10\% & 0.0093 & 0.0103 & \textbf{0.0110} & 18.28\% & 6.80\% \\
    \multicolumn{1}{c|}{} & N@20 & 0.0091 & 0.0134 & \textbf{0.0153} & 68.13\% & 14.18\% & 0.0142 & 0.0157 & \textbf{0.0170} & 19.72\% & 8.28\% & 0.0096 & 0.0105 & \textbf{0.0116} & 20.83\% & 10.48\% & 0.0129 & 0.0139 & \textbf{0.0146} & 13.18\% & 5.04\% \\
    \midrule
    
    \multicolumn{1}{c|}{\multirow{6}[0]{*}{Yelp}} & H@5 & 0.0092 & 0.0119 & \textbf{0.0149} & 61.96\% & 25.21\% & 0.0140 & 0.0159 & \textbf{0.0164} & 17.14\% & 3.14\% & 0.0131 & 0.0144 & \textbf{0.0148} & 12.98\% & 2.78\% & 0.0109 & 0.0121 & \textbf{0.0128} & 17.43\% & 5.79\% \\
    \multicolumn{1}{c|}{} & H@10 & 0.0172 & 0.0223 & \textbf{0.0274} & 59.30\% & 22.87\% & 0.0255 & 0.0277 & \textbf{0.0296} & 16.08\% & 6.86\% & 0.0238 & 0.0252 & \textbf{0.0269} & 13.03\% & 6.75\% & 0.0218 & 0.0240 & \textbf{0.0252} & 15.60\% & 5.00\% \\
    \multicolumn{1}{c|}{} & H@20 & 0.0314 & 0.0407 & \textbf{0.0474} & 50.96\% & 16.46\% & 0.0443 & 0.0482 & \textbf{0.0505} & 14.00\% & 4.77\% & 0.0414 & 0.0435 & \textbf{0.0503} & 21.50\% & 15.63\% & 0.0400 & 0.0433 & \textbf{0.0451} & 12.75\% & 4.16\% \\
    \multicolumn{1}{c|}{} & N@5 & 0.0057 & 0.0072 & \textbf{0.0089} & 56.14\% & 23.61\% & 0.0084 & 0.0098 & \textbf{0.0100} & 19.05\% & 2.04\% & 0.0084 & \textbf{0.0089} & 0.0087 & 3.57\% & -2.25\% & 0.0075 & 0.0081 & \textbf{0.0086} & 14.67\% & 6.17\% \\
    \multicolumn{1}{c|}{} & N@10 & 0.0083 & 0.0105 & \textbf{0.0129} & 55.42\% & 22.86\% & 0.0120 & 0.0136 & \textbf{0.0142} & 18.33\% & 4.41\% & 0.0118 & 0.0127 & \textbf{0.0134} & 13.56\% & 5.51\% & 0.0106 & 0.0117 & \textbf{0.0125} & 17.92\% & 6.84\% \\
    \multicolumn{1}{c|}{} & N@20 & 0.0118 & 0.0151 & \textbf{0.0179} & 51.69\% & 18.54\% & 0.0167 & 0.0188 & \textbf{0.0194} & 16.17\% & 3.19\% & 0.0162 & 0.0173 & \textbf{0.0184} & 13.58\% & 6.36\% & 0.0149 & 0.0162 & \textbf{0.0166} & 11.41\% & 2.47\% \\
    \midrule
    
    \multicolumn{1}{c|}{\multirow{6}[0]{*}{Home}} & H@5 & 0.0030 & 0.0063 & \textbf{0.0073} & 143.33\% & 15.87\% & 0.0059 & 0.0073 & \textbf{0.0078} & 32.20\% & 6.85\% & 0.0035 & 0.0044 & \textbf{0.0047} & 34.29\% & 6.82\% & 0.0039 & 0.0043 & \textbf{0.0046} & 17.95\% & 6.98\% \\
    \multicolumn{1}{c|}{} & H@10 & 0.0057 & 0.0108 & \textbf{0.0121} & 112.28\% & 12.04\% & 0.0109 & 0.0125 & \textbf{0.0135} & 23.85\% & 8.00\% & 0.0066 & 0.0075 & \textbf{0.0080} & 21.21\% & 6.67\% & 0.0074 & 0.0085 & \textbf{0.0089} & 20.27\% & 4.71\% \\
    \multicolumn{1}{c|}{} & H@20 & 0.0106 & 0.0182 & \textbf{0.0201} & 89.62\% & 10.44\% & 0.0185 & 0.0205 & \textbf{0.0216} & 16.76\% & 5.37\% & 0.0125 & 0.0136 & \textbf{0.0146} & 16.80\% & 7.35\% & 0.0137 & 0.0151 & \textbf{0.0160} & 16.79\% & 5.96\% \\
    \multicolumn{1}{c|}{} & N@5 & 0.0018 & 0.0040 & \textbf{0.0047} & 161.11\% & 17.50\% & 0.0037 & 0.0048 & \textbf{0.0052} & 40.54\% & 8.33\% & 0.0022 & 0.0026 & \textbf{0.0029} & 31.82\% & 11.54\% & 0.0025 & 0.0029 & \textbf{0.0032} & 28.00\% & 10.34\% \\
    \multicolumn{1}{c|}{} & N@10 & 0.0027 & 0.0054 & \textbf{0.0063} & 133.33\% & 16.67\% & 0.0053 & 0.0065 & \textbf{0.0070} & 32.08\% & 7.69\% & 0.0031 & 0.0039 & \textbf{0.0040} & 29.03\% & 2.56\% & 0.0037 & 0.0042 & \textbf{0.0045} & 21.62\% & 7.14\% \\
    \multicolumn{1}{c|}{} & N@20 & 0.0039 & 0.0073 & \textbf{0.0083} & 112.82\% & 13.70\% & 0.0072 & 0.0085 & \textbf{0.0090} & 25.00\% & 5.88\% & 0.0046 & 0.0050 & \textbf{0.0056} & 21.74\% & 12.00\% & 0.0051 & 0.0059 & \textbf{0.0063} & 23.53\% & 6.78\% \\
    \midrule
    
    \multicolumn{2}{c|}{Average} & - & - & - & 84.11\% & 14.85\% & - & - & - & 20.86\% & 8.38\% & - & - & - & 24.24\% & 8.17\% & - & - & - & 19.03\% & 5.57\% \\
    
    \midrule
    \multicolumn{22}{c}{} \\
    \midrule
    \multicolumn{1}{c|}{Dataset} & Metric & SASRec & w/ RP & w/ RP+ & Imp-O & Imp-R & LightSANs & w/ RP & w/ RP+ & Imp-O & Imp-R & gMLP & w/ RP & w/ RP+ & Imp-O & Imp-R & FMLP-Rec & w/ RP & w/ RP+ & Imp-O & Imp-R \\
    \midrule
    
    \multicolumn{1}{c|}{\multirow{6}[0]{*}{Toys}} & H@5 & 0.0472 & 0.0587 & \textbf{0.0622} & 31.78\% & 5.96\% & 0.0376 & 0.0388 & \textbf{0.0438} & 16.49\% & 12.89\% & 0.0373 & 0.0385 & \textbf{0.0394} & 5.63\% & 2.34\% & 0.0418 & 0.0463 & \textbf{0.0473} & 13.16\% & 2.16\% \\
    \multicolumn{1}{c|}{} & H@10 & 0.0716 & 0.0843 & \textbf{0.0866} & 20.95\% & 2.73\% & 0.0505 & 0.0564 & \textbf{0.0614} & 21.58\% & 8.87\% & 0.0495 & 0.0508 & \textbf{0.0523} & 5.66\% & 2.95\% & 0.0570 & 0.0632 & \textbf{0.0660} & 15.79\% & 4.43\% \\
    \multicolumn{1}{c|}{} & H@20 & 0.0965 & 0.1142 & \textbf{0.1172} & 21.45\% & 2.63\% & 0.0656 & 0.0770 & \textbf{0.0843} & 28.51\% & 9.48\% & 0.0665 & 0.0685 & \textbf{0.0712} & 7.07\% & 3.94\% & 0.0759 & 0.0852 & \textbf{0.0899} & 18.45\% & 5.52\% \\
    \multicolumn{1}{c|}{} & N@5 & 0.0317 & 0.0393 & \textbf{0.0433} & 36.59\% & 10.18\% & 0.0265 & 0.0260 & \textbf{0.0308} & 16.23\% & 18.46\% & 0.0269 & 0.0276 & \textbf{0.0285} & 5.95\% & 3.26\% & 0.0289 & 0.0314 & \textbf{0.0333} & 15.22\% & 6.05\% \\
    \multicolumn{1}{c|}{} & N@10 & 0.0396 & 0.0476 & \textbf{0.0512} & 29.29\% & 7.56\% & 0.0307 & 0.0316 & \textbf{0.0364} & 18.57\% & 15.19\% & 0.0308 & 0.0321 & \textbf{0.0333} & 8.12\% & 3.74\% & 0.0338 & 0.0368 & \textbf{0.0393} & 16.27\% & 6.79\% \\
    \multicolumn{1}{c|}{} & N@20 & 0.0471 & 0.0551 & \textbf{0.0589} & 25.05\% & 6.90\% & 0.0345 & 0.0368 & \textbf{0.0422} & 22.32\% & 14.67\% & 0.0351 & 0.0359 & \textbf{0.0378} & 7.69\% & 5.29\% & 0.0386 & 0.0423 & \textbf{0.0454} & 17.62\% & 7.33\% \\
    \midrule
    
    \multicolumn{1}{c|}{\multirow{6}[0]{*}{Beauty}} & H@5 & 0.0409 & 0.0477 & \textbf{0.0512} & 25.18\% & 7.34\% & 0.0312 & 0.0328 & \textbf{0.0398} & 27.56\% & 21.34\% & 0.0313 & \textbf{0.0361} & 0.0355 & 13.42\% & -1.66\% & 0.0387 & 0.0439 & \textbf{0.0459} & 18.60\% & 4.56\% \\
    \multicolumn{1}{c|}{} & H@10 & 0.0637 & 0.0726 & \textbf{0.0762} & 19.62\% & 4.96\% & 0.0469 & 0.0485 & \textbf{0.0598} & 27.51\% & 23.30\% & 0.0486 & 0.0525 & \textbf{0.0539} & 10.91\% & 2.67\% & 0.0566 & 0.0648 & \textbf{0.0664} & 17.31\% & 2.47\% \\
    \multicolumn{1}{c|}{} & H@20 & 0.0942 & 0.1034 & \textbf{0.1089} & 15.61\% & 5.32\% & 0.0709 & 0.0741 & \textbf{0.0854} & 20.45\% & 15.25\% & 0.0703 & 0.0748 & \textbf{0.0765} & 8.82\% & 2.27\% & 0.0814 & 0.0905 & \textbf{0.0935} & 14.86\% & 3.31\% \\
    \multicolumn{1}{c|}{} & N@5 & 0.0257 & 0.0310 & \textbf{0.0344} & 33.85\% & 10.97\% & 0.0201 & 0.0217 & \textbf{0.0269} & 33.83\% & 23.96\% & 0.0209 & 0.0239 & \textbf{0.0244} & 16.75\% & 2.09\% & 0.0260 & 0.0286 & \textbf{0.0305} & 17.31\% & 6.64\% \\
    \multicolumn{1}{c|}{} & N@10 & 0.0331 & 0.0390 & \textbf{0.0424} & 28.10\% & 8.72\% & 0.0255 & 0.0267 & \textbf{0.0334} & 30.98\% & 25.09\% & 0.0264 & 0.0292 & \textbf{0.0314} & 18.94\% & 7.53\% & 0.0317 & 0.0353 & \textbf{0.0371} & 17.03\% & 5.10\% \\
    \multicolumn{1}{c|}{} & N@20 & 0.0408 & 0.0468 & \textbf{0.0507} & 24.26\% & 8.33\% & 0.0319 & 0.0323 & \textbf{0.0398} & 24.76\% & 23.22\% & 0.0319 & 0.0348 & \textbf{0.0361} & 13.17\% & 3.74\% & 0.0380 & 0.0418 & \textbf{0.0439} & 15.53\% & 5.02\% \\
    \midrule
    
    \multicolumn{1}{c|}{\multirow{6}[0]{*}{Sports}} & H@5 & 0.0187 & 0.0237 & \textbf{0.0294} & 57.22\% & 24.05\% & 0.0130 & 0.0150 & \textbf{0.0173} & 33.08\% & 15.33\% & 0.0136 & 0.0152 & \textbf{0.0169} & 24.26\% & 11.18\% & 0.0149 & 0.0201 & \textbf{0.0209} & 40.27\% & 3.98\% \\
    \multicolumn{1}{c|}{} & H@10 & 0.0299 & 0.0369 & \textbf{0.0434} & 45.15\% & 17.62\% & 0.0208 & 0.0231 & \textbf{0.0280} & 34.62\% & 21.21\% & 0.0210 & 0.0229 & \textbf{0.0248} & 18.10\% & 8.30\% & 0.0238 & 0.0294 & \textbf{0.0307} & 28.99\% & 4.42\% \\
    \multicolumn{1}{c|}{} & H@20 & 0.0447 & 0.0553 & \textbf{0.0641} & 43.40\% & 15.91\% & 0.0327 & 0.0352 & \textbf{0.0417} & 27.52\% & 18.47\% & 0.0328 & 0.0342 & \textbf{0.0365} & 11.28\% & 6.73\% & 0.0345 & 0.0439 & \textbf{0.0481} & 39.42\% & 9.57\% \\
    \multicolumn{1}{c|}{} & N@5 & 0.0124 & 0.0154 & \textbf{0.0194} & 56.45\% & 25.97\% & 0.0087 & 0.0097 & \textbf{0.0114} & 31.03\% & 17.53\% & 0.0090 & 0.0101 & \textbf{0.0112} & 24.44\% & 10.89\% & 0.0097 & \textbf{0.0132} & 0.0129 & 32.99\% & -2.27\% \\
    \multicolumn{1}{c|}{} & N@10 & 0.0160 & 0.0196 & \textbf{0.0239} & 49.38\% & 21.94\% & 0.0112 & 0.0122 & \textbf{0.0148} & 32.14\% & 21.31\% & 0.0114 & 0.0126 & \textbf{0.0137} & 20.18\% & 8.73\% & 0.0126 & \textbf{0.0162} & 0.0157 & 24.60\% & -3.09\% \\
    \multicolumn{1}{c|}{} & N@20 & 0.0197 & 0.0242 & \textbf{0.0291} & 47.72\% & 20.25\% & 0.0142 & 0.0153 & \textbf{0.0182} & 28.17\% & 18.95\% & 0.0143 & 0.0154 & \textbf{0.0167} & 16.78\% & 8.44\% & 0.0153 & 0.0199 & \textbf{0.0206} & 34.64\% & 3.52\% \\
    \midrule
    
    \multicolumn{1}{c|}{\multirow{6}[0]{*}{Yelp}} & H@5 & 0.0168 & 0.0210 & \textbf{0.0219} & 30.36\% & 4.29\% & 0.0120 & 0.0159 & \textbf{0.0163} & 35.83\% & 2.52\% & 0.0108 & 0.0115 & \textbf{0.0140} & 29.63\% & 21.74\% & 0.0121 & 0.0154 & \textbf{0.0213} & 76.03\% & 38.31\% \\
    \multicolumn{1}{c|}{} & H@10 & 0.0282 & 0.0359 & \textbf{0.0376} & 33.33\% & 4.74\% & 0.0207 & 0.0267 & \textbf{0.0280} & 35.27\% & 4.87\% & 0.0190 & 0.0201 & \textbf{0.0239} & 25.79\% & 18.91\% & 0.0209 & 0.0258 & \textbf{0.0319} & 52.63\% & 23.64\% \\
    \multicolumn{1}{c|}{} & H@20 & 0.0443 & 0.0587 & \textbf{0.0598} & 34.99\% & 1.87\% & 0.0346 & 0.0429 & \textbf{0.0437} & 26.30\% & 1.86\% & 0.0325 & 0.0338 & \textbf{0.0416} & 28.00\% & 23.08\% & 0.0356 & 0.0425 & \textbf{0.0467} & 31.18\% & 9.88\% \\
    \multicolumn{1}{c|}{} & N@5 & 0.0106 & 0.0129 & \textbf{0.0136} & 28.30\% & 5.43\% & 0.0073 & 0.0099 & \textbf{0.0103} & 41.10\% & 4.04\% & 0.0066 & 0.0072 & \textbf{0.0085} & 28.79\% & 18.06\% & 0.0070 & 0.0099 & \textbf{0.0144} & 105.71\% & 45.45\% \\
    \multicolumn{1}{c|}{} & N@10 & 0.0142 & 0.0177 & \textbf{0.0188} & 32.39\% & 6.21\% & 0.0101 & 0.0134 & \textbf{0.0139} & 37.62\% & 3.73\% & 0.0093 & 0.0099 & \textbf{0.0116} & 24.73\% & 17.17\% & 0.0098 & 0.0133 & \textbf{0.0178} & 81.63\% & 33.83\% \\
    \multicolumn{1}{c|}{} & N@20 & 0.0183 & 0.0234 & \textbf{0.0247} & 34.97\% & 5.56\% & 0.0136 & 0.0175 & \textbf{0.0180} & 32.35\% & 2.86\% & 0.0126 & 0.0134 & \textbf{0.0161} & 27.78\% & 20.15\% & 0.0135 & 0.0175 & \textbf{0.0215} & 59.26\% & 22.86\% \\
    \midrule
    
    \multicolumn{1}{c|}{\multirow{6}[0]{*}{Home}} & H@5 & 0.0096 & 0.0130 & \textbf{0.0136} & 41.67\% & 4.62\% & 0.0059 & 0.0065 & \textbf{0.0085} & 44.07\% & 30.77\% & 0.0057 & 0.0080 & \textbf{0.0083} & 45.61\% & 3.75\% & 0.0079 & 0.0085 & \textbf{0.0110} & 39.24\% & 29.41\% \\
    \multicolumn{1}{c|}{} & H@10 & 0.0155 & 0.0204 & \textbf{0.0211} & 36.13\% & 3.43\% & 0.0100 & 0.0098 & \textbf{0.0133} & 33.00\% & 35.71\% & 0.0092 & 0.0123 & \textbf{0.0126} & 36.96\% & 2.44\% & 0.0125 & 0.0128 & \textbf{0.0165} & 32.00\% & 28.91\% \\
    \multicolumn{1}{c|}{} & H@20 & 0.0244 & 0.0311 & \textbf{0.0324} & 32.79\% & 4.18\% & 0.0173 & 0.0185 & \textbf{0.0202} & 16.76\% & 9.19\% & 0.0157 & 0.0181 & \textbf{0.0193} & 22.93\% & 6.63\% & 0.0193 & 0.0191 & \textbf{0.0243} & 25.91\% & 27.23\% \\
    \multicolumn{1}{c|}{} & N@5 & 0.0061 & 0.0087 & \textbf{0.0090} & 47.54\% & 3.45\% & 0.0034 & 0.0039 & \textbf{0.0057} & 67.65\% & 46.15\% & 0.0037 & 0.0054 & \textbf{0.0057} & 54.05\% & 5.56\% & 0.0051 & 0.0055 & \textbf{0.0092} & 80.39\% & 67.27\% \\
    \multicolumn{1}{c|}{} & N@10 & 0.0083 & 0.0116 & \textbf{0.0124} & 49.40\% & 6.90\% & 0.0047 & 0.0050 & \textbf{0.0072} & 53.19\% & 44.00\% & 0.0049 & 0.0067 & \textbf{0.0071} & 44.90\% & 5.97\% & 0.0066 & 0.0069 & \textbf{0.0090} & 36.36\% & 30.43\% \\
    \multicolumn{1}{c|}{} & N@20 & 0.0101 & 0.0137 & \textbf{0.0145} & 43.56\% & 5.84\% & 0.0064 & 0.0067 & \textbf{0.0089} & 39.06\% & 32.84\% & 0.0065 & 0.0082 & \textbf{0.0086} & 32.31\% & 4.88\% & 0.0083 & 0.0085 & \textbf{0.0109} & 31.33\% & 28.24\% \\
    \midrule

    \multicolumn{2}{c|}{Average} & - & - & - & 35.34\% & 8.92\% & - & - & - & 31.25\% & 18.10\% & - & - & - & 21.29\% & 8.03\% & - & - & - & 34.99\% & 15.37\% \\
    \bottomrule
    \end{tabular}}}
  \label{tab:main_short}%
\end{table}%

\begin{table}[!ht]
  \centering
  \caption{Performance improvement on two long-sequence datasets. The `w/ RP' and `w/ RP+' represent adding RepPad and RepPad+ to the corresponding model, respectively. The best performance of the three is bolded. The `Imp-O' and `Imp-R' represent the relative improvements of RepPad+ over the original model and the RepPad, respectively.}
  \renewcommand\arraystretch{1.0}
  \setlength{\tabcolsep}{1mm}{
  \scalebox{0.59}{
    \begin{tabular}{cc|ccccc|ccccc|ccccc|ccccc}
    \toprule
    \multicolumn{1}{c|}{Dataset} & Metric & GRU4Rec & w/ RP & w/ RP+ & Imp-O & Imp-R & NARM & w/ RP & w/ RP+ & Imp-O & Imp-R & Caser & Imp-OA & Imp-RP & Imp-O & Imp-R & NextItNet & w/ RP & w/ RP+ & Imp-O & Imp-R \\
    \midrule
    
    \multicolumn{1}{c|}{\multirow{6}[0]{*}{LastFM}} & H@5 & 0.0229 & 0.0330 & \textbf{0.0353} & 54.15\% & 6.97\% & 0.0505 & 0.0578 & \textbf{0.0596} & 18.02\% & 3.11\% & 0.0321 & 0.0394 & \textbf{0.0426} & 32.71\% & 8.12\% & 0.0312 & 0.0339 & \textbf{0.0367} & 17.63\% & 8.26\% \\
    \multicolumn{1}{c|}{} & H@10 & 0.0339 & 0.0541 & \textbf{0.0582} & 71.68\% & 7.58\% & 0.0716 & 0.0862 & \textbf{0.0890} & 24.30\% & 3.25\% & 0.0550 & 0.0679 & \textbf{0.0707} & 28.55\% & 4.12\% & 0.0560 & 0.0565 & \textbf{0.0679} & 21.25\% & 20.18\% \\
    \multicolumn{1}{c|}{} & H@20 & 0.0624 & 0.0706 & \textbf{0.0869} & 39.26\% & 23.09\% & 0.1037 & 0.1220 & \textbf{0.1312} & 26.52\% & 7.54\% & 0.0862 & 0.0908 & \textbf{0.1055} & 22.39\% & 16.19\% & 0.0855 & 0.0927 & \textbf{0.1138} & 33.10\% & 22.76\% \\
    \multicolumn{1}{c|}{} & N@5 & 0.0152 & 0.0225 & \textbf{0.0243} & 59.87\% & 8.00\% & 0.0341 & 0.0374 & \textbf{0.0400} & 17.30\% & 6.95\% & 0.0226 & 0.0292 & \textbf{0.0308} & 36.28\% & 5.48\% & 0.0192 & 0.0215 & \textbf{0.0243} & 26.56\% & 13.02\% \\
    \multicolumn{1}{c|}{} & N@10 & 0.0188 & 0.0292 & \textbf{0.0307} & 63.30\% & 5.14\% & 0.0408 & 0.0465 & \textbf{0.0494} & 21.08\% & 6.24\% & 0.0300 & 0.0383 & \textbf{0.0396} & 32.00\% & 3.39\% & 0.0272 & 0.0285 & \textbf{0.0341} & 25.37\% & 19.65\% \\
    \multicolumn{1}{c|}{} & N@20 & 0.0259 & 0.0333 & \textbf{0.0385} & 48.65\% & 15.62\% & 0.0490 & 0.0556 & \textbf{0.0599} & 22.24\% & 7.73\% & 0.0380 & 0.0442 & \textbf{0.0470} & 23.68\% & 6.33\% & 0.0348 & 0.0376 & \textbf{0.0457} & 31.32\% & 21.54\% \\
    \midrule
    
    \multicolumn{1}{c|}{\multirow{6}[0]{*}{ML-1M}} & H@5 & 0.1104 & 0.1407 & \textbf{0.1492} & 35.14\% & 6.04\% & 0.0493 & \textbf{0.0541} & 0.0532 & 7.91\% & -1.66\% & 0.0465 & 0.0495 & \textbf{0.0507} & 9.03\% & 2.42\% & 0.0922 & 0.1008 & \textbf{0.1052} & 14.10\% & 4.37\% \\
    \multicolumn{1}{c|}{} & H@10 & 0.1841 & 0.2200 & \textbf{0.2313} & 25.64\% & 5.14\% & 0.0868 & 0.0879 & \textbf{0.0923} & 6.34\% & 5.01\% & 0.0823 & 0.0841 & \textbf{0.0862} & 4.74\% & 2.50\% & 0.1491 & 0.1623 & \textbf{0.1674} & 12.27\% & 3.14\% \\
    \multicolumn{1}{c|}{} & H@20 & 0.2849 & 0.3300 & \textbf{0.3482} & 22.22\% & 5.52\% & 0.1459 & 0.1470 & \textbf{0.1548} & 6.10\% & 5.31\% & 0.1416 & 0.1429 & \textbf{0.1501} & 6.00\% & 5.04\% & 0.2354 & 0.2519 & \textbf{0.2600} & 10.45\% & 3.22\% \\
    \multicolumn{1}{c|}{} & N@5 & 0.0704 & 0.0914 & \textbf{0.0942} & 33.81\% & 3.06\% & 0.0297 & 0.0349 & \textbf{0.0360} & 21.21\% & 3.15\% & 0.0277 & 0.0304 & \textbf{0.0310} & 11.91\% & 1.97\% & 0.0620 & 0.0643 & \textbf{0.0657} & 5.97\% & 2.18\% \\
    \multicolumn{1}{c|}{} & N@10 & 0.0941 & 0.1168 & \textbf{0.1207} & 28.27\% & 3.34\% & 0.0417 & 0.0457 & \textbf{0.0464} & 11.27\% & 1.53\% & 0.0398 & 0.0414 & \textbf{0.0426} & 7.04\% & 2.90\% & 0.0815 & 0.0866 & \textbf{0.0883} & 8.34\% & 1.96\% \\
    \multicolumn{1}{c|}{} & N@20 & 0.1195 & 0.1446 & \textbf{0.1502} & 25.69\% & 3.87\% & 0.0565 & 0.0606 & \textbf{0.0628} & 11.15\% & 3.63\% & 0.0541 & 0.0562 & \textbf{0.0572} & 5.73\% & 1.78\% & 0.0963 & 0.1025 & \textbf{0.1063} & 10.38\% & 3.71\% \\
    \midrule
    \multicolumn{2}{c|}{Average} & - & - & - & 42.31\% & 7.78\% & - & - & - & 16.12\% & 4.32\% & - & - & - & 18.34\% & 5.02\% & - & - & - & 18.06\% & 10.33\% \\
    \midrule
    \multicolumn{22}{c}{} \\
    \midrule
    
    \multicolumn{1}{c|}{Dataset} & Metric & SASRec & w/ RP & w/ RP+ & Imp-O & Imp-R & LightSANs & w/ RP & w/ RP+ & Imp-O & Imp-R & gMLP & w/ RP & w/ RP+ & Imp-O & Imp-R & FMLP-Rec & w/ RP & w/ RP+ & Imp-O & Imp-R \\
    \midrule
    
    \multicolumn{1}{c|}{\multirow{6}[0]{*}{LastFM}} & H@5 & 0.0440 & 0.0505 & \textbf{0.0569} & 29.32\% & 12.67\% & \textbf{0.0422} & 0.0303 & 0.0385 & -8.77\% & 27.06\% & 0.0239 & 0.0247 & \textbf{0.0349} & 46.03\% & 41.30\% & 0.0257 & 0.0321 & \textbf{0.0376} & 46.30\% & 17.13\% \\
    \multicolumn{1}{c|}{} & H@10 & 0.0606 & 0.0789 & \textbf{0.0826} & 36.30\% & 4.69\% & 0.0560 & 0.0560 & \textbf{0.0679} & 21.25\% & 21.25\% & 0.0385 & 0.0422 & \textbf{0.0541} & 40.52\% & 28.20\% & 0.0431 & 0.0459 & \textbf{0.0532} & 23.43\% & 15.90\% \\
    \multicolumn{1}{c|}{} & H@20 & 0.0972 & 0.1220 & \textbf{0.1284} & 32.10\% & 5.25\% & 0.0789 & 0.0817 & \textbf{0.0936} & 18.63\% & 14.57\% & 0.0670 & 0.0695 & \textbf{0.0862} & 28.66\% & 24.03\% & 0.0560 & 0.0661 & \textbf{0.0752} & 34.29\% & 13.77\% \\
    \multicolumn{1}{c|}{} & N@5 & 0.0287 & 0.0336 & \textbf{0.0373} & 29.97\% & 11.01\% & \textbf{0.0274} & 0.0186 & 0.0233 & -14.96\% & 25.27\% & 0.0140 & 0.0150 & \textbf{0.0241} & 72.14\% & 60.67\% & 0.0170 & 0.0208 & \textbf{0.0214} & 25.88\% & 2.88\% \\
    \multicolumn{1}{c|}{} & N@10 & 0.0340 & 0.0430 & \textbf{0.0454} & 33.53\% & 5.58\% & 0.0318 & 0.0267 & \textbf{0.0327} & 2.83\% & 22.47\% & 0.0186 & 0.0217 & \textbf{0.0304} & 63.44\% & 40.09\% & 0.0224 & 0.0253 & \textbf{0.0259} & 15.63\% & 2.37\% \\
    \multicolumn{1}{c|}{} & N@20 & 0.0434 & 0.0538 & \textbf{0.0567} & 30.65\% & 5.39\% & 0.0377 & 0.0333 & \textbf{0.0393} & 4.24\% & 18.02\% & 0.0257 & 0.0275 & \textbf{0.0384} & 49.42\% & 39.64\% & 0.0256 & 0.0304 & \textbf{0.0315} & 23.05\% & 3.62\% \\
    \midrule
    
    \multicolumn{1}{c|}{\multirow{6}[0]{*}{ML-1M}} & H@5 & 0.1748 & 0.1674 & \textbf{0.1818} & 4.00\% & 8.60\% & 0.1179 & 0.1096 & \textbf{0.1295} & 9.84\% & 18.16\% & 0.0601 & 0.0742 & \textbf{0.0868} & 44.43\% & 16.98\% & 0.1045 & 0.1109 & \textbf{0.1328} & 27.08\% & 19.75\% \\
    \multicolumn{1}{c|}{} & H@10 & 0.2624 & 0.2608 & \textbf{0.2715} & 3.47\% & 4.10\% & 0.1873 & 0.1758 & \textbf{0.2026} & 8.17\% & 15.24\% & 0.1070 & 0.1180 & \textbf{0.1323} & 23.64\% & 12.12\% & 0.1667 & 0.1639 & \textbf{0.1851} & 11.04\% & 12.93\% \\
    \multicolumn{1}{c|}{} & H@20 & 0.3762 & 0.3780 & \textbf{0.3839} & 2.05\% & 1.56\% & 0.2765 & 0.2647 & \textbf{0.3003} & 8.61\% & 13.45\% & 0.1717 & 0.1882 & \textbf{0.1998} & 16.37\% & 6.16\% & 0.2454 & 0.2445 & \textbf{0.2609} & 6.32\% & 6.71\% \\
    \multicolumn{1}{c|}{} & N@5 & 0.1171 & 0.1096 & \textbf{0.1173} & 0.17\% & 7.03\% & 0.0770 & 0.0712 & \textbf{0.0842} & 9.35\% & 18.26\% & 0.0368 & 0.0483 & \textbf{0.0551} & 49.73\% & 14.08\% & 0.0670 & 0.0719 & \textbf{0.0829} & 23.73\% & 15.30\% \\
    \multicolumn{1}{c|}{} & N@10 & 0.1450 & 0.1396 & \textbf{0.1462} & 0.83\% & 4.73\% & 0.0993 & 0.0924 & \textbf{0.1078} & 8.56\% & 16.67\% & 0.0518 & 0.0625 & \textbf{0.0697} & 34.56\% & 11.52\% & 0.0871 & 0.0889 & \textbf{0.1025} & 17.68\% & 15.30\% \\
    \multicolumn{1}{c|}{} & N@20 & 0.1737 & 0.1693 & \textbf{0.1745} & 0.46\% & 3.07\% & 0.1218 & 0.1153 & \textbf{0.1323} & 8.62\% & 14.74\% & 0.0680 & 0.0800 & \textbf{0.0867} & 27.50\% & 8.38\% & 0.1069 & 0.1092 & \textbf{0.1216} & 13.75\% & 11.36\% \\
    \midrule
    
    \multicolumn{2}{c|}{Average} & - & - & - & 16.90\% & 6.14\% & - & - & - & 6.36\% & 18.76\% & - & - & - & 41.37\% & 25.26\%& - & - & - & 22.35\% & 11.42\% \\
    \bottomrule
    \end{tabular}}}
  \label{tab:main_long}%
\end{table}%

\subsection{Overall Improvements}
The experimental results of original sequential recommendation models and adding our method are presented in Table \ref{tab:main_short} and Table \ref{tab:main_long}. These tables show that our RepPad+ can significantly improve the performance of sequential recommendation models. The improvement on the long-sequence datasets is less overall than on the short-sequence datasets. These results validate our conjecture in Section \ref{sec:Limitations}. More specifically, we have the following observations: 

(1) As a whole, the RNN- and CNN-based models perform comparably. NARM is usually the best performer in both categories, which verifies the effectiveness of combining attention mechanisms and RNN architectures for sequential recommendation. The Transformer-based models perform better overall, especially SASRec. This shows the ability of the Transformer architecture to model user sequence patterns. MLP-Based models perform slightly less well, which may be related to the fact that the pure MLP architecture cannot accurately capture the fine-grained relationship between each user interaction and historical behavior as the Transformer does. 

(2) As demonstrated in Table \ref{tab:main_short}, after adding our RepPad+ to the original model, the performance of all models based on different architectures is significantly improved. Overall, RepPad+ brings more performance improvements to the Transformer and RNN-based models than the MLP and CNN-based ones. For the most representative and commonly used recommendation models, the average performance improvement is up to 84.11\% on GRU4Rec and 35.38\% on SASRec. The above results validate the effectiveness of our method on short-sequence datasets. It also shows that RepPad+ can be applied to various types of sequential models. Considering that in real-world scenarios, the vast majority of user history records are short sequences \cite{liu2021augmenting, jiang2021sequential}, our approach has great potential for application. For RepPad and RepPad+, We observe that the latter still delivers sizable performance gains over the former. This phenomenon suggests that medium-length sequences are still a part of the dataset, and augmentation of them is necessary.

(3) From Table \ref{tab:main_long}, we observe that RepPad+ still achieves competitive improvements on long-sequence datasets. In the previous work by Dang et al. \cite{dang2024repeated}, RepPad not only fails to improve model performance when faced with long-sequence datasets but even causes performance degradation on some methods. In the long-sequence dataset, excluding the sequences whose length exceeds the maximum sequence length $N$, a large portion of the user sequences are medium-length sequences that cannot satisfy the full repeated padding, RepPad+ takes into account the augmentation of these medium-length sequences, and thus achieves a significant performance improvement compared to RepPad. This result shows that our method can be applied not only to different kinds of sequential models but also to datasets with different sparsity levels. From the perspective of different models, The SASRec still outperforms other models overall, further demonstrating the power of Transformer architecture.

\begin{table}[!t]
 \centering
    \caption{Performance comparison between previous work RepPad, our method RepPad+, and heuristic data augmentation methods. The best performance is bolded and the second best is underlined.}
    \renewcommand\arraystretch{1.0}
   \setlength{\tabcolsep}{1.25mm}{
  \scalebox{0.62}{
    \begin{tabular}{c|c|c|ccccc|cccc|c|c|ccccc|cc}
\cmidrule{1-10}\cmidrule{12-21} \multicolumn{10}{c}{GRU4Rec} & \qquad & \multicolumn{10}{c}{SASRec} \\
\cmidrule{1-10}\cmidrule{12-21} Dataset & Metric & Original & Ran & SW & Ran-S & CMR & CMRSI & RepPad & RepPad+ & & Dataset & Metric & Original & Ran & SW & Ran-S & CMR & CMRSI & RepPad & RepPad+ \\
\cmidrule{1-10}\cmidrule{12-21} \multicolumn{1}{c|}{\multirow{4}[2]{*}{Toys}} & H@10 & 0.0256 & 0.0306 & 0.0342 & 0.0363 & 0.0361 & 0.0411 & \underline{0.0434} & \textbf{0.0482} & & \multicolumn{1}{c|}{\multirow{4}[2]{*}{Toys}} & H@10 & 0.0716 & 0.0600 & 0.0522 & 0.0660 & 0.0634 & 0.0713 & \underline{0.0843} & \textbf{0.0866} \\
        & H@20 & 0.0432 & 0.0491 & 0.0557 & 0.0594 & 0.0585 & 0.0629 & \underline{0.0652} & \textbf{0.0716} &    &    & H@20 & 0.0965 & 0.0875 & 0.0807 & 0.0950 & 0.0952 & 0.1039 & \underline{0.1142} & \textbf{0.1172} \\
       & N@10 & 0.0121 & 0.0158 & 0.0169 & 0.0192 & 0.0181 & 0.0203 & \underline{0.0229} & \textbf{0.0269} &     &     & N@10 & 0.0396 & 0.0321 & 0.0292 & 0.0366 & 0.0347 & 0.0400 & \underline{0.0476} & \textbf{0.0512} \\
        & N@20 & 0.0166 & 0.0205 & 0.0223 & 0.0248 & 0.0238 & 0.0267 & \underline{0.0284} & \textbf{0.0328} &     &     & N@20 & 0.0471 & 0.0390 & 0.0363 & 0.0439 & 0.0428 & 0.0482 & \underline{0.0551} & \textbf{0.0589} \\
\cmidrule{1-10}\cmidrule{12-21}    \multirow{4}[2]{*}{Beauty} & H@10 & 0.0327 & 0.0370 & 0.0365 & 0.0419 & 0.0398 & 0.0463 & \underline{0.0468} & \textbf{0.0505} &     & \multirow{4}[2]{*}{Beauty} & H@10 & 0.0637 & 0.0553 & 0.0542 & 0.0563 & 0.0562 & 0.0603 & \underline{0.0726} & \textbf{0.0762} \\
        & H@20 & 0.0551 & 0.0581 & 0.0604 & 0.0650 & 0.0673 & 0.0712 & \underline{0.0732} & \textbf{0.0756} &     &     & H@20 & 0.0942 & 0.0850 & 0.0846 & 0.0882 & 0.0876 & 0.0940 & \underline{0.1034} & \textbf{0.1089} \\
        & N@10 & 0.0162 & 0.0191 & 0.0176 & 0.0203 & 0.0196 & 0.0224 & \underline{0.0234} & \textbf{0.0263} &     &     & N@10 & 0.0331 & 0.0285 & 0.0270 & 0.0289 & 0.0290 & 0.0316 & \underline{0.0390} & \textbf{0.0424} \\
        & N@20 & 0.0219 & 0.0241 & 0.0236 & 0.0255 & 0.0266 & 0.0290 & \underline{0.0300} & \textbf{0.0326} &     &     & N@20 & 0.0408 & 0.0359 & 0.0346 & 0.0369 & 0.0639 & 0.0401 & \underline{0.0468} & \textbf{0.0507} \\
\cmidrule{1-10}\cmidrule{12-21} \multirow{4}[2]{*}{Sports} & H@10 & 0.0143 & 0.0163 & 0.0165 & 0.0194 & 0.0186 & 0.0176 & \underline{0.0204} & \textbf{0.0232} & & \multirow{4}[2]{*}{Sports} & H@10 & 0.0299 & 0.0330 & 0.0315 & 0.0308 & 0.0332 & 0.0318 & \underline{0.0369} & \textbf{0.0434} \\
        & H@20 & 0.0249 & 0.0267 & 0.0301 & 0.0319 & 0.0309 & 0.0296 & \underline{0.0338} & \textbf{0.0383} &     &     & H@20 & 0.0447 & 0.0520 & 0.0496 & 0.0514 & 0.0530 & 0.0504 & \underline{0.0553} & \textbf{0.0641} \\
        & N@10 & 0.0065 & 0.0080 & 0.0078 & 0.0096 & 0.0093 & 0.0088 & \underline{0.0100} & \textbf{0.0115} &     &     & N@10 & 0.0160 & 0.0177 & 0.0165 & 0.0154 & 0.0173 & 0.0173 & \underline{0.0196} & \textbf{0.0239} \\
         & N@20 & 0.0091 & 0.0106 & 0.0112 & 0.0127 & 0.0124 & 0.0117 & \underline{0.0134} & \textbf{0.0153} &      &      & N@20 & 0.0197 & 0.0225 & 0.0211 & 0.0206 & 0.0223 & 0.0212 & \underline{0.0242} & \textbf{0.0291} \\
\cmidrule{1-10}\cmidrule{12-21} \multirow{4}[2]{*}{Yelp} & H@10 & 0.0172 & 0.0184 & 0.0195 & 0.0210 & 0.0203 & 0.0214 & \underline{0.0223} & \textbf{0.0274} &      & \multirow{4}[2]{*}{Yelp} & H@10 & 0.0282 & 0.0316 & 0.0279 & \underline{0.0364} & 0.0278 & 0.0310 & 0.0359 & \textbf{0.0374} \\
         & H@20 & 0.0314 & 0.0347 & 0.0349 & 0.0380 & 0.0375 & 0.0389 & \underline{0.0407} & \textbf{0.0474} &      &      & H@20 & 0.0443 & 0.0516 & 0.0483 & \textbf{0.0620} & 0.0480 & 0.0523 & 0.0587 & \underline{0.0601} \\
         & N@10 & 0.0083 & 0.0085 & 0.0090 & 0.0096 & 0.0103 & 0.0101 & \underline{0.0105} & \textbf{0.0129} &      &      & N@10 & 0.0142 & 0.0162 & 0.0137 & \underline{0.0184} & 0.0136 & 0.0156 & 0.0177 & \textbf{0.0190} \\
         & N@20 & 0.0118 & 0.0126 & 0.0133 & 0.0142 & 0.0139 & 0.0147 & \underline{0.0151} & \textbf{0.0179} &      &      & N@20 & 0.0183 & 0.0212 & 0.0189 & \underline{0.0248} & 0.0187 & 0.0210 & 0.0234 & \textbf{0.0255} \\
\cmidrule{1-10}\cmidrule{12-21}    \multirow{4}[2]{*}{Home} & H@10 & 0.0057 & 0.0064 & 0.0072 & 0.0087 & 0.0078 & 0.0094 & \underline{0.0108} & \textbf{0.0121} &      & \multirow{4}[2]{*}{Home} & H@10 & 0.0155 & 0.0156 & 0.0166 & \underline{0.0208} & 0.0167 & 0.0173 & 0.0204 & \textbf{0.0211} \\
         & H@20 & 0.0106 & 0.0120 & 0.0141 & 0.0151 & 0.0149 & 0.0159 & \underline{0.0182} & \textbf{0.0201} &      &      & H@20 & 0.0244 & 0.0247 & 0.0270 & \textbf{0.0336} & 0.0275 & 0.0276 & 0.0311 & \underline{0.0324} \\
         & N@10 & 0.0027 & 0.0030 & 0.0034 & 0.0042 & 0.0037 & 0.0044 & \underline{0.0054} & \textbf{0.0063} &      &      & N@10 & 0.0083 & 0.0086 & 0.0089 & 0.0109 & 0.0092 & 0.0099 & \underline{0.0110} & \textbf{0.0124} \\
         & N@20 & 0.0039 & 0.0045 & 0.0051 & 0.0059 & 0.0054 & 0.0060 & \underline{0.0073} & \textbf{0.0083} &      &      & N@20 & 0.0101 & 0.0105 & 0.0111 & \underline{0.0141} & 0.0113 & 0.0122 & 0.0137 & \textbf{0.0145} \\
\cmidrule{1-10}\cmidrule{12-21}    
\end{tabular}}}%
\label{tab:reppad_vs_heuristic}%
\end{table}%

\subsection{Compare with Other Methods}
In this subsection, we compare RepPad+ with different data augmentation methods. For the heuristic methods, we choose the two representative models, SASRec and GRU4Rec, as the backbone network. For methods that require training, we chose the most versatile SASRec since different methods have different restrictions on the type of backbone network. For example, the ASReP is based on Transformer architecture and can not be applied to GRU4Rec. We present the results in Table \ref{tab:reppad_vs_heuristic} and \ref{tab:reppad_vs_train}.

\begin{table}[!ht]
  \centering
  \caption{Performance comparison between previous work RepPad, our method RepPad+, and training-required data augmentation methods. The best performance is bolded and the second best is underlined. The backbone network is SASRec.}
  \scalebox{0.78}{
    \begin{tabular}{c|c|c|ccc|cc}
    \toprule
    Dataset & Metric & Original & ASReP & DiffuASR & CL4SRec & RepPad & RepPad+ \\
    \midrule
    \multicolumn{1}{c|}{\multirow{4}[1]{*}{Toys}} & H@10 & 0.0716 & 0.0755 & 0.0783 & 0.0804 & \underline{0.0843} & \textbf{0.0866} \\
         & H@20 & 0.0965 & 0.1021 & 0.1085 & 0.1052 & \underline{0.1142} & \textbf{0.1172} \\
         & N@10 & 0.0396 & 0.0423 & 0.0453 & 0.0447 & \underline{0.0476} & \textbf{0.0512} \\
         & N@20 & 0.0471 & 0.0506 & 0.0535 & 0.0520 & \underline{0.0551} & \textbf{0.0589} \\
    \midrule
    \multirow{4}[0]{*}{Beauty} & H@10 & 0.0637 & 0.0664 & 0.0679 & 0.0686 & \underline{0.0726} & \textbf{0.0762} \\
         & H@20 & 0.0942 & 0.0959 & 0.0966 & 0.0982 & \underline{0.1034} & \textbf{0.1089} \\
         & N@10 & 0.0331 & 0.0351 & 0.0372 & 0.0366 & \underline{0.0390} & \textbf{0.0424} \\
         & N@20 & 0.0408 & 0.0422 & 0.0454 & 0.0449 & \underline{0.0468} & \textbf{0.0507} \\
    \midrule
    \multirow{4}[0]{*}{Sports} & H@10 & 0.0299 & 0.0324 & \underline{0.0375} & 0.0339 & 0.0369 & \textbf{0.0434} \\
         & H@20 & 0.0447 & 0.0503 & \underline{0.0566} & 0.0521 & 0.0553 & \textbf{0.0641} \\
         & N@10 & 0.0160 & 0.0166 & 0.0169 & 0.0172 & \underline{0.0196} & \textbf{0.0239} \\
         & N@20 & 0.0197 & 0.0201 & 0.0214 & 0.0225 & \underline{0.0242} & \textbf{0.0291} \\
    \midrule
    \multirow{4}[0]{*}{Yelp} & H@10 & 0.0282 & 0.0323 & 0.0345 & \textbf{0.0385} & 0.0359 & \underline{0.0374}  \\
         & H@20 & 0.0443 & 0.0539 & 0.0566 & \textbf{0.0621} & 0.0587 & \underline{0.0601}  \\
         & N@10 & 0.0142 & 0.0153 & 0.0169 & \underline{0.0182} & 0.0177 & \textbf{0.0190} \\
         & N@20 & 0.0183 & 0.0207 & 0.0228 & \underline{0.0251} & 0.0234 & \textbf{0.0255} \\
    \midrule
    \multirow{4}[0]{*}{Home} & H@10 & 0.0155 & 0.0176 & 0.0198 & \textbf{0.0220} & 0.0204 & \underline{0.0211}  \\
         & H@20 & 0.0244 & 0.0285 & 0.0301 & \textbf{0.0329} & 0.0311 & \underline{0.0324}  \\
         & N@10 & 0.0083 & 0.0098 & 0.0113 & \underline{0.0122} & 0.0116 & \textbf{0.0124} \\
         & N@20 & 0.0101 & 0.0121 & 0.0124 & \textbf{0.0149} & 0.0137 & \underline{0.0145}  \\
    \bottomrule
    \end{tabular}%
  \label{tab:reppad_vs_train}}%
\end{table}%

From Table \ref{tab:reppad_vs_heuristic}, we observe that RepPad+ outperforms existing heuristic data augmentation methods and the previous work RepPad. Among the baseline methods, CMRSI usually performs the best, thanks to the five well-designed data augmentation operators that help mine more preference information while mitigating data sparsity. Ran-S also achieves competitive performance in some examples due to its augmentation using items in the original sequence. In some cases, the model performance degraded after using heuristic data augmentation methods, which may be related to the fact that these methods introduce too much noise that destroys the integrity of the original sequence. In most cases, RepPad can achieve the second-best performance, while RepPad+ can achieve the best. From Table \ref{tab:reppad_vs_train}, RepPad+ still achieves satisfactory results in the face of data augmentation methods that require training. Among the baseline methods, DiffuASR and CL4SRec leverage the diffusion model and sequence-level augmentation with contrastive learning to achieve competitive performance in some cases. Surprisingly, RepPad+ achieves the best or second-best performance in all cases without including any parameters and training process. 

Overall, RepPad+ proved its superiority by achieving highly competitive results without including any parameters or training processes. Compared to RepPad, RepPad+ also demonstrated the importance of augmenting the medium-length sequences. Also, our approach is not limited by the backbone network and can be seamlessly inserted into most existing sequential recommendation models.

\subsection{Improvement on Long and Short Sequences}
Following BiCAT \cite{jiang2021sequential}, we explore the effectiveness of our methods on short and long sequences in the same dataset and report the results in Table \ref{tab:different_length_GRU4Rec} and \ref{tab:different_length_SASRec}. A long sequence is a sequence whose length $\left|s_u\right|$ fulfill the condition $20 < \left|s_u\right|$. Since we perform 5-core filtering during the dataset, a short sequence is an interaction sequence of length 5 (i.e., $5 = \left|s_u\right|$). From the tables, we observe a similar trend as in Table \ref{tab:main_long}. With the addition of the RepPad, model performance can often achieve significant gains on short sequences while achieving limited performance gains on long sequences, and in many cases, even producing sizable performance degradation. Compared to RepPad, RepPad+ improves the model's performance on long sequences by augmenting medium-length sequences. For items in long sequences, the model can learn their representations more accurately, and these items are likely to appear in short sequences, so RepPad+ improves the model's recommendation performance for short sequences as well. Overall, RepPad+ makes up for RepPad's poor performance on long sequences and further improves the overall recommendation accuracy of the model. Comparing the results of our method on long and short sequences, our method is highly applicable to mitigating the cold start problem, as well as facilitating the learning of long-tailed item representations.

\begin{table}[!t]
  \centering
  \caption{Model performance on different sequences. The `w/ RP' and `w/ RP+' represent adding RepPad and our RepPad+ to the corresponding model, respectively. The `Imp-O' and `Imp-R' represent the relative improvements of RepPad+ over the original model and the RepPad, respectively. The backbone network is GRU4Rec.}
  \scalebox{0.70}{
    \begin{tabular}{cc|c|cc|ccc|ccc|ccc}
    \toprule
    \multicolumn{2}{c|}{Sequence Length} & \multicolumn{6}{c|}{Long ($20 < \left|s_u\right| \leq 50$)} & \multicolumn{6}{c}{Short ($\left|s_u\right| = 5$)} \\
    \midrule
    \multicolumn{1}{c|}{Dataset} & Metric & Original & w/ RP & Imp-O & w/ RP+ & Imp-O & Imp-R & Original & w/ RP & Imp-O & w/ RP+ & Imp-O & Imp-R \\
    \midrule
    \multicolumn{1}{c|}{\multirow{4}[0]{*}{Toys}} & H@10 & 0.0314 & 0.0392 & 24.84\% & 0.0597 & 90.13\% & 52.30\% & 0.0251 & 0.0354 & 41.04\% & 0.0419 & 66.93\% & 18.36\% \\
    \multicolumn{1}{c|}{} & H@20 & 0.0769 & 0.0689 & -10.40\% & 0.1052 & 36.80\% & 52.69\% & 0.0426 & 0.0534 & 25.35\% & 0.0642 & 50.70\% & 20.22\% \\
    \multicolumn{1}{c|}{} & N@10 & 0.0133 & 0.0207 & 55.64\% & 0.0279 & 109.77\% & 34.78\% & 0.0130 & 0.0199 & 53.08\% & 0.0222 & 70.77\% & 11.56\% \\
    \multicolumn{1}{c|}{} & N@20 & 0.0245 & 0.0273 & 11.61\% & 0.0391 & 59.85\% & 43.22\% & 0.0174 & 0.0244 & 40.23\% & 0.0278 & 59.77\% & 13.93\% \\
    \midrule
    \multicolumn{1}{c|}{\multirow{4}[0]{*}{Beauty}} & H@10 & 0.0663 & 0.0889 & 34.09\% & 0.1139 & 71.79\% & 28.12\% & 0.0264 & 0.0350 & 32.58\% & 0.0489 & 85.23\% & 39.71\% \\
    \multicolumn{1}{c|}{} & H@20 & 0.1189 & 0.1377 & 15.81\% & 0.1564 & 31.54\% & 13.58\% & 0.0417 & 0.0588 & 41.01\% & 0.0711 & 70.50\% & 20.92\% \\
    \multicolumn{1}{c|}{} & N@10 & 0.0383 & 0.0497 & 29.77\% & 0.0587 & 53.26\% & 18.11\% & 0.0124 & 0.0172 & 38.71\% & 0.0293 & 136.29\% & 70.35\% \\
    \multicolumn{1}{c|}{} & N@20 & 0.0516 & 0.0618 & 19.77\% & 0.0695 & 34.69\% & 12.46\% & 0.0163 & 0.0231 & 41.72\% & 0.0349 & 114.11\% & 51.08\% \\
    \midrule
    \multicolumn{1}{c|}{\multirow{4}[0]{*}{Sports}} & H@10 & 0.0212 & 0.0193 & -8.96\% & 0.0198 & -6.60\% & 2.59\% & 0.0138 & 0.0203 & 47.10\% & 0.0259 & 87.68\% & 27.59\% \\
    \multicolumn{1}{c|}{} & H@20 & 0.0308 & 0.0285 & -7.47\% & 0.0353 & 14.61\% & 23.86\% & 0.0243 & 0.0348 & 43.21\% & 0.0414 & 70.37\% & 18.97\% \\
    \multicolumn{1}{c|}{} & N@10 & 0.0097 & 0.0088 & -9.28\% & 0.0093 & -4.12\% & 5.68\% & 0.0063 & 0.0102 & 61.90\% & 0.0128 & 103.17\% & 25.49\% \\
    \multicolumn{1}{c|}{} & N@20 & 0.0122 & 0.0114 & -6.56\% & 0.0132 & 8.20\% & 15.79\% & 0.0089 & 0.0138 & 55.06\% & 0.0167 & 87.64\% & 21.01\% \\
    \midrule
    \multicolumn{1}{c|}{\multirow{4}[0]{*}{Yelp}} & H@10 & 0.0160 & 0.0176 & 10.00\% & 0.0267 & 66.88\% & 51.70\% & 0.0176 & 0.0235 & 33.52\% & 0.0255 & 44.89\% & 8.51\% \\
    \multicolumn{1}{c|}{} & H@20 & 0.0342 & 0.0374 & 9.36\% & 0.0465 & 35.96\% & 24.33\% & 0.0332 & 0.0410 & 23.49\% & 0.0435 & 31.02\% & 6.10\% \\
    \multicolumn{1}{c|}{} & N@10 & 0.0081 & 0.0082 & 1.23\% & 0.0115 & 41.98\% & 40.24\% & 0.0085 & 0.0114 & 34.12\% & 0.0125 & 47.06\% & 9.65\% \\
    \multicolumn{1}{c|}{} & N@20 & 0.0126 & 0.0131 & 3.97\% & 0.0166 & 31.75\% & 26.72\% & 0.0124 & 0.0161 & 29.84\% & 0.0170 & 37.10\% & 5.59\% \\
    \midrule
    \multicolumn{1}{c|}{\multirow{4}[0]{*}{Home}} & H@10 & 0.0241 & 0.0213 & -11.62\% & 0.0257 & 6.64\% & 20.66\% & 0.0042 & 0.0089 & 111.90\% & 0.0116 & 176.19\% & 30.34\% \\
    \multicolumn{1}{c|}{} & H@20 & 0.0408 & 0.0354 & -13.24\% & 0.0389 & -4.66\% & 9.89\% & 0.0076 & 0.0150 & 97.37\% & 0.0185 & 143.42\% & 23.33\% \\
    \multicolumn{1}{c|}{} & N@10 & 0.0115 & 0.0107 & -6.96\% & 0.0123 & 6.96\% & 14.95\% & 0.0020 & 0.0048 & 140.00\% & 0.0062 & 210.00\% & 29.17\% \\
    \multicolumn{1}{c|}{} & N@20 & 0.0158 & 0.0142 & -10.13\% & 0.0157 & -0.63\% & 10.56\% & 0.0029 & 0.0064 & 120.69\% & 0.0079 & 172.41\% & 23.44\% \\
    \midrule
    \multicolumn{2}{c|}{Average} & - & - & 6.57\% & - & 34.24\% & 25.11\% & - & - & 55.60\% & - & 93.26\% & 23.77\% \\
    \bottomrule
    \end{tabular}}%
  \label{tab:different_length_GRU4Rec}%
\end{table}%

\begin{table}[!t]
  \centering
  \caption{Model performance on different sequences. The `w/ RP' and `w/ RP+' represent adding RepPad and our RepPad+ to the corresponding model, respectively. The `Imp-O' and `Imp-R' represent the relative improvements of RepPad+ over the original model and the RepPad, respectively. The backbone network is SASRec.}
  \scalebox{0.70}{
    \begin{tabular}{cc|c|cc|ccc|c|cc|ccc}
    \toprule
    \multicolumn{2}{c|}{Sequence Length} & \multicolumn{6}{c|}{Long ($20 < \left|s_u\right| \leq 50$)} & \multicolumn{6}{c}{Short ($\left|s_u\right| = 5$)} \\
    \midrule
    \multicolumn{1}{c|}{Dataset} & Metric & Original & w/ RP & Imp-O & w/ RP+ & Imp-O & Imp-R & Original & w/ RP & Imp-O & w/ RP+ & Imp-O & Imp-R \\
    \midrule
    \multicolumn{1}{c|}{\multirow{4}[0]{*}{Toys}} & H@10 & 0.0863 & 0.0863 & 0.00\% & 0.0973 & 12.75\% & 12.75\% & 0.0743 & 0.0807 & 8.61\% & 0.0888 & 19.52\% & 10.04\% \\
    \multicolumn{1}{c|}{} & H@20 & 0.1334 & 0.1303 & -2.32\% & 0.1381 & 3.52\% & 5.99\% & 0.0974 & 0.1075 & 10.37\% & 0.1161 & 19.20\% & 8.00\% \\
    \multicolumn{1}{c|}{} & N@10 & 0.0474 & 0.0400 & -15.61\% & 0.0546 & 15.19\% & 36.50\% & 0.0435 & 0.0472 & 8.51\% & 0.0543 & 24.83\% & 15.04\% \\
    \multicolumn{1}{c|}{} & N@20 & 0.0594 & 0.0511 & -13.97\% & 0.0650 & 9.43\% & 27.20\% & 0.0493 & 0.0539 & 9.33\% & 0.0612 & 24.14\% & 13.54\% \\
    \midrule
    \multicolumn{1}{c|}{\multirow{4}[0]{*}{Beauty}} & H@10 & 0.1250 & 0.1261 & 0.88\% & 0.1327 & 6.16\% & 5.23\% & 0.0526 & 0.0623 & 18.44\% & 0.0702 & 33.46\% & 12.68\% \\
    \multicolumn{1}{c|}{} & H@20 & 0.1779 & 0.1711 & -3.82\% & 0.1902 & 6.91\% & 11.16\% & 0.0797 & 0.0881 & 10.54\% & 0.0987 & 23.84\% & 12.03\% \\
    \multicolumn{1}{c|}{} & N@10 & 0.0654 & 0.0693 & 5.96\% & 0.0773 & 18.20\% & 11.54\% & 0.0280 & 0.0346 & 23.57\% & 0.0381 & 36.07\% & 10.12\% \\
    \multicolumn{1}{c|}{} & N@20 & 0.0788 & 0.0806 & 2.28\% & 0.0919 & 16.62\% & 14.02\% & 0.0348 & 0.0412 & 18.39\% & 0.0452 & 29.89\% & 9.71\% \\
    \midrule
    \multicolumn{1}{c|}{\multirow{4}[0]{*}{Sports}} & H@10 & 0.0408 & 0.0397 & -2.70\% & 0.0441 & 8.09\% & 11.08\% & 0.0328 & 0.0396 & 20.73\% & 0.0469 & 42.99\% & 18.43\% \\
    \multicolumn{1}{c|}{} & H@20 & 0.0540 & 0.0507 & -6.11\% & 0.0606 & 12.22\% & 19.53\% & 0.0510 & 0.0587 & 15.10\% & 0.0706 & 38.43\% & 20.27\% \\
    \multicolumn{1}{c|}{} & N@10 & 0.0224 & 0.0223 & -0.45\% & 0.0241 & 7.59\% & 8.07\% & 0.0167 & 0.0215 & 28.74\% & 0.0260 & 55.69\% & 20.93\% \\
    \multicolumn{1}{c|}{} & N@20 & 0.0258 & 0.0250 & -3.10\% & 0.0282 & 9.30\% & 12.80\% & 0.0213 & 0.0263 & 23.47\% & 0.0320 & 50.23\% & 21.67\% \\
    \midrule
    \multicolumn{1}{c|}{\multirow{4}[0]{*}{Yelp}} & H@10 & 0.0271 & 0.0262 & -3.32\% & 0.0347 & 28.04\% & 32.44\% & 0.0230 & 0.0318 & 38.26\% & 0.0349 & 51.74\% & 9.75\% \\
    \multicolumn{1}{c|}{} & H@20 & 0.0474 & 0.0459 & -3.16\% & 0.0545 & 14.98\% & 18.74\% & 0.0347 & 0.0540 & 55.62\% & 0.0599 & 72.62\% & 10.93\% \\
    \multicolumn{1}{c|}{} & N@10 & 0.0136 & 0.0137 & 0.74\% & 0.0166 & 22.06\% & 21.17\% & 0.0106 & 0.0154 & 45.28\% & 0.0172 & 62.26\% & 11.69\% \\
    \multicolumn{1}{c|}{} & N@20 & 0.0186 & 0.0188 & 1.08\% & 0.0215 & 15.59\% & 14.36\% & 0.0136 & 0.0210 & 54.41\% & 0.0235 & 72.79\% & 11.90\% \\
    \midrule
    \multicolumn{1}{c|}{\multirow{4}[0]{*}{Home}} & H@10 & 0.0374 & 0.0277 & -25.94\% & 0.0344 & -8.02\% & 24.19\% & 0.0122 & 0.0195 & 59.84\% & 0.0208 & 70.49\% & 6.67\% \\
    \multicolumn{1}{c|}{} & H@20 & 0.0505 & 0.0479 & -5.15\% & 0.0521 & 3.17\% & 8.77\% & 0.0188 & 0.0292 & 55.32\% & 0.0323 & 71.81\% & 10.62\% \\
    \multicolumn{1}{c|}{} & N@10 & 0.0189 & 0.0137 & -27.51\% & 0.0178 & -5.82\% & 29.93\% & 0.0065 & 0.0109 & 67.69\% & 0.0117 & 80.00\% & 7.34\% \\
    \multicolumn{1}{c|}{} & N@20 & 0.0222 & 0.0209 & -5.86\% & 0.0223 & 0.45\% & 6.70\% & 0.0080 & 0.0134 & 67.50\% & 0.0146 & 82.50\% & 8.96\% \\
    \midrule
    \multicolumn{2}{c|}{Average} & - & - & -5.40\% & - & 9.82\% & 16.61\% & - & - & 31.99\% & - & 48.12\% & 12.52\% \\
    \bottomrule
    \end{tabular}}%
  \label{tab:different_length_SASRec}%
\end{table}%

\subsection{Ablation Study}
\begin{figure}[!t]
	\centering
	\includegraphics[scale=0.47]{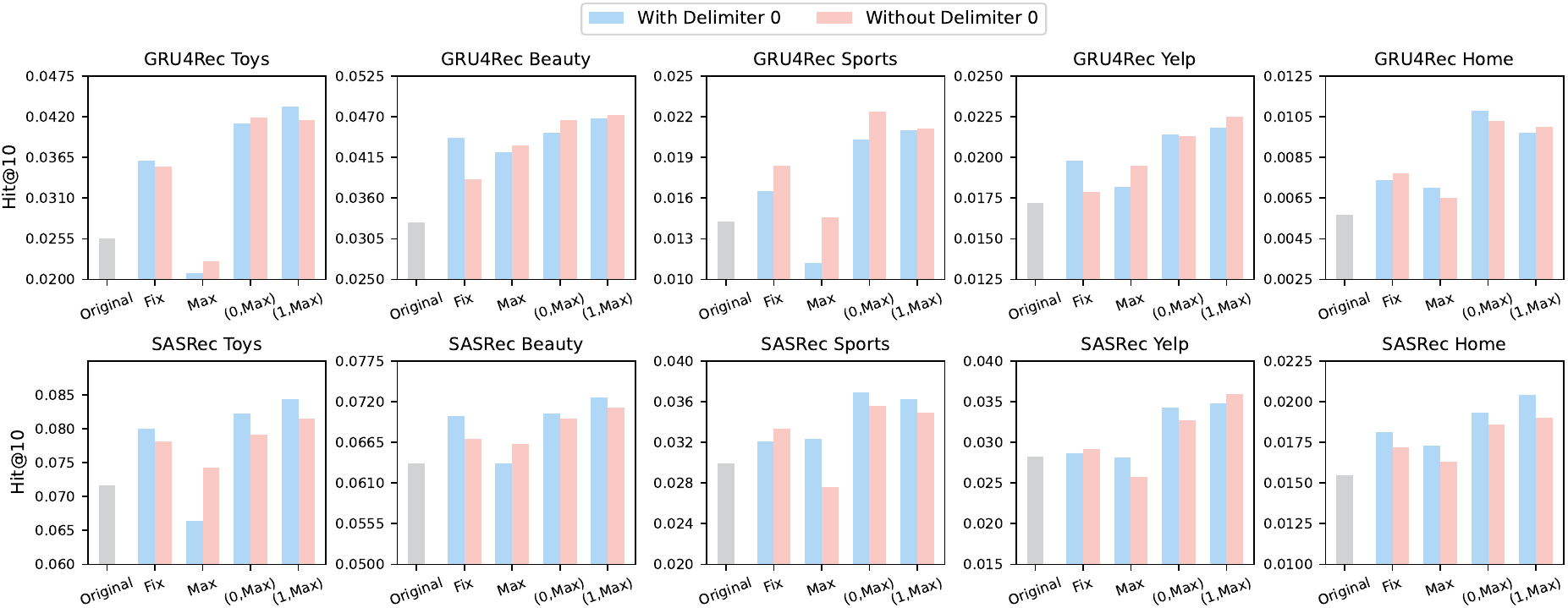}
	\caption{The result of the ablation study for RepPad.}
	\label{fig:variant_1}
\end{figure}

\begin{figure}[!t]
	\centering
	\includegraphics[scale=0.47]{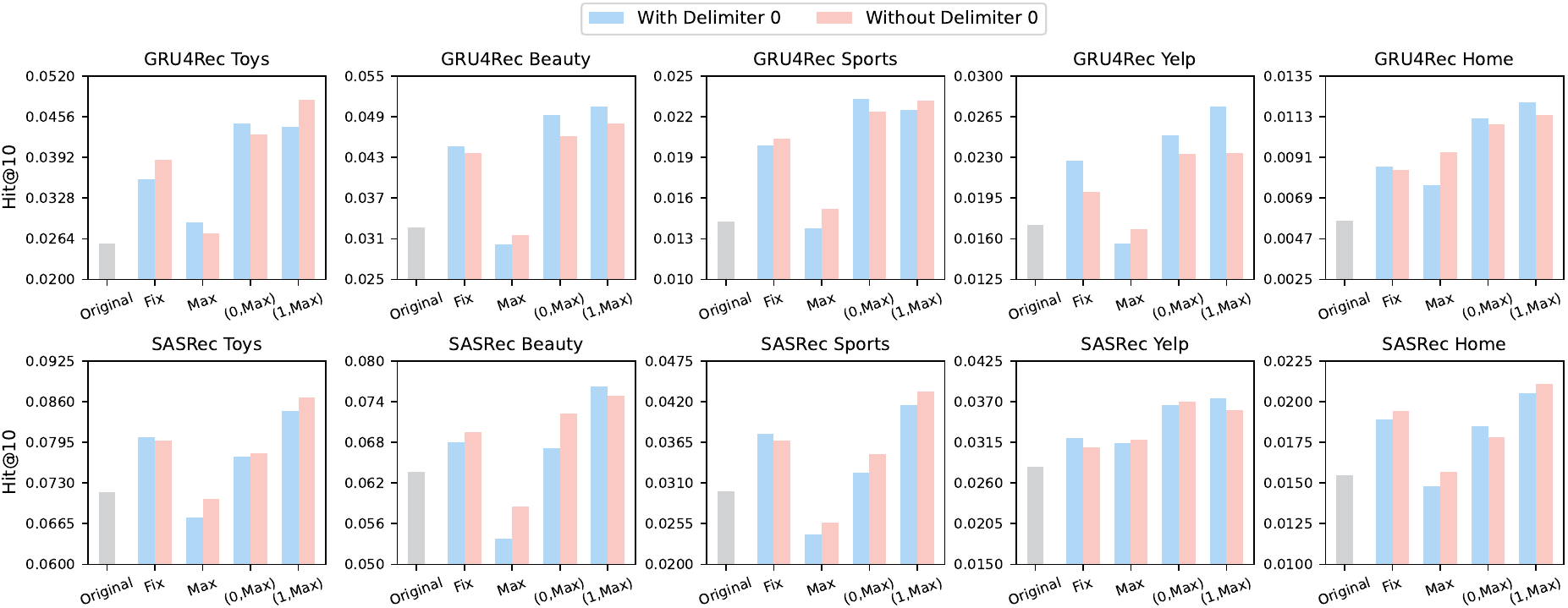}
	\caption{The result of the ablation study for RepPad+.}
	\label{fig:variant_2}
\end{figure}

We conduct an ablation study to explore the different variants of RepPad and RepPad+. Based on the padding times $m$ in Table \ref{tab:valueofm} and whether or not to add the delimiter \emph{0}, we compare the recommendation performance of the following variants: 1) \textbf{Fix}: Performing repeated padding with a fixed number of times, we report the best results for times in range $\{1, 2, 3\}$. 2) \textbf{Max}: Performing repeated padding until the remaining space is insufficient (i.e. maximum padding times). 3) \textbf{(0, Max)}: When performing repeated padding each time, randomly select between 0 and the maximum number of times. 4) \textbf{(1, Max)}: When performing repeated padding each time, randomly select between 1 and the maximum number of times. Figure \ref{fig:variant_1} and \ref{fig:variant_2} presents the variants and results. We report the performance of each of the four variants with and without the addition of the delimiter \emph{0}.

Compared to the original model, RepPad+ delivers more overall performance gains than RepPad. Looking at the different variants, Fix and Max bring limited performance improvement compared to (0, Max) and (1, Max). The performance of (0, Max) and (1, Max) can come with considerable improvements. Further, (1, Max) performs slightly better than (0, Max) on the whole, indicating that performing repeat padding for short sequences at least once is essential. For Fix, our methods are performed with a fixed number of times, so it does not fully utilize the idle input space. For Max, we observe that it not only does not bring improvement in some cases but also leads to a decrease in model performance. If every sequence performs repeated padding to the maximum number of times, it may make the representation learned by the model unbalanced. Because the repeated padding time tends to be large for short sequences, multiple iterations of short behavioral sequences may interfere with the model's ability for long-sequence and long-term preferences modeling, which in turn impairs recommendation performance. For (0, Max) and (1, Max), they avoid the imbalance problem caused by RP-Max while using the idle input space wisely. 

For whether or not to add the delimiter \emph{0}, we observe that this factor is related to the type of recommendation model in Figure \ref{fig:variant_1}. For GRU4Rec, there are more examples of better performance when no delimiter is added. In contrast, for SASRec, there are more examples that models perform better when the delimiter is added. In other words, RNN-based models are less susceptible to the interference of "predicting the tail with the head". The self-attention module of Transformer-based models pays attention to the effect of each historical interaction on the following action, so the negative effect of "predicting the tail with the head" can be weakened by adding the delimiter \emph{0}. In Figure \ref{fig:variant_2}, these trends are just the opposite, i.e., GRU4Rec performs better with the addition of delimiter \emph{0} and SASRec performs better without the delimiter \emph{0}. In addition, we observe that the need to add the delimiter \emph{0} is also relevant to the dataset. For example, for variant (1, Max), the trend of performance advantages and disadvantages with and without \emph{0} is very similar when using different backbone networks under the same dataset. As the only adjustable option in our method, we recommend users try adding \emph{0} and not adding it respectively in real applications.

\begin{table}[!t]
  \centering
  \caption{The training speed (seconds) of RepPad+ and different baseline methods. We give the increased time compared to the original model. The best performance is bolded, and the second best is underlined.}
 \renewcommand\arraystretch{1.0}
   \scalebox{0.85}{
    \begin{tabular}{c|c|ccc|c|ccc}
    \toprule
    \multirow{2}[4]{*}{Dataset} & \multicolumn{4}{c|}{GRU4Rec} & \multicolumn{4}{c}{SASRec} \\
\cmidrule{2-9}        & Method & Speed (s/epoch) & Total Time (s) & H@10 & Method & Speed (s/epoch) & Total Time (s) & H@10 \\
    \midrule
    \multirow{5}[4]{*}{Beauty} & Original & 0.960 & 153 & 0.0327 & Original & 2.389 & 395 & 0.0637 \\
\cmidrule{2-9}        & w/ CMRSI & 2.547 (+1.587) & 507 (+354) & \underline{0.0463} & w/ CMRSI & 4.561 (+2.172) & 1416 (+1021) & 0.0603 \\
        & w/ CL4SRec & 1.748 (+0.778) & 329 (+176) & 0.0440 & w/ CL4SRec & 3.954 (+1.565) & 677 (+282) & 0.0686 \\
        & w/ Fix & \textbf{1.103 (+0.143)} & \underline{203 (+60)} & 0.0447 & w/ Fix & \textbf{2.574 (+0.185)} & \underline{522 (+127)} & \underline{0.0695}  \\
        & w/ (1,Max) & \underline{1.392 (+0.432)} & \textbf{168 (+15)} & \textbf{0.0505 } & w/ (1,Max) & \underline{2.806 (0.0417)} & \textbf{503 (+108)} & \textbf{0.0762} \\
    \midrule
    \multirow{5}[4]{*}{Sports} & Original & 1.422 & 210 & 0.0143 & Original & 3.133 & 642 & 0.0299 \\
\cmidrule{2-9}        & w/ CMRSI & 3.853 (+2.431) & 1081 (+871) & 0.0176 & w/ CMRSI & 6.400 (+3.267) & 1066 (+424) & 0.0318 \\
        & w/ CL4SRec & 2.620 (+1.198) & 561 (+351) & \textbf{0.0271 } & w/ CL4SRec & 5.479 (+2.346) & 928 (+286) & 0.0339 \\
        & w/ Fix & \textbf{1.585 (+0.163)} & \underline{248 (+38)} & 0.0204 & w/ Fix & \textbf{3.237 (+0.104)} & \textbf{746 (+104)} & \underline{0.0376} \\
        & w/ (1,Max) & \underline{2.087 (+0.655)} & \textbf{261 (+51)} & \underline{0.0232} & w/ (1,Max) & \underline{3.782 (+0.649)} & \underline{861 (+219)} & \textbf{0.0434} \\
    \midrule
    \multirow{5}[4]{*}{Home} & Original & 2.531 & 391 & 0.0057 & Original & 4.840 & 958 & 0.0155 \\
\cmidrule{2-9}        & w/ CMRSI & 16.013 (+13.482) & 1560 (+1169) & 0.0094 & w/ CMRSI & 13.464 (+8.624) & 2837 (+1879) & 0.0173 \\
        & w/ CL4SRec & 5.101 (+2.570) & 939 (+548) & \textbf{0.0156 } & w/ CL4SRec & 9.196 (+4.356) & 1938 (+980) & \textbf{0.0220 } \\
        & w/ Fix & \textbf{2.786 (+0.255)} & \textbf{400 (+9)} & 0.0086 & w/ Fix & \textbf{5.156 (+0.316)} & \textbf{1077 (+119)} & 0.0189 \\
        & w/ (1,Max) & \underline{3.745 (+1.214)} & \underline{421 (+30)} & \underline{0.0121} & w/ (1,Max) & \underline{5.987 (+1.147)} & \underline{1147 (+189)} & \underline{0.0211} \\
    \bottomrule
    \end{tabular}%
  \label{tab:speed}}%
\end{table}%

\begin{figure}[!t]
	\centering
	\includegraphics[scale=0.65]{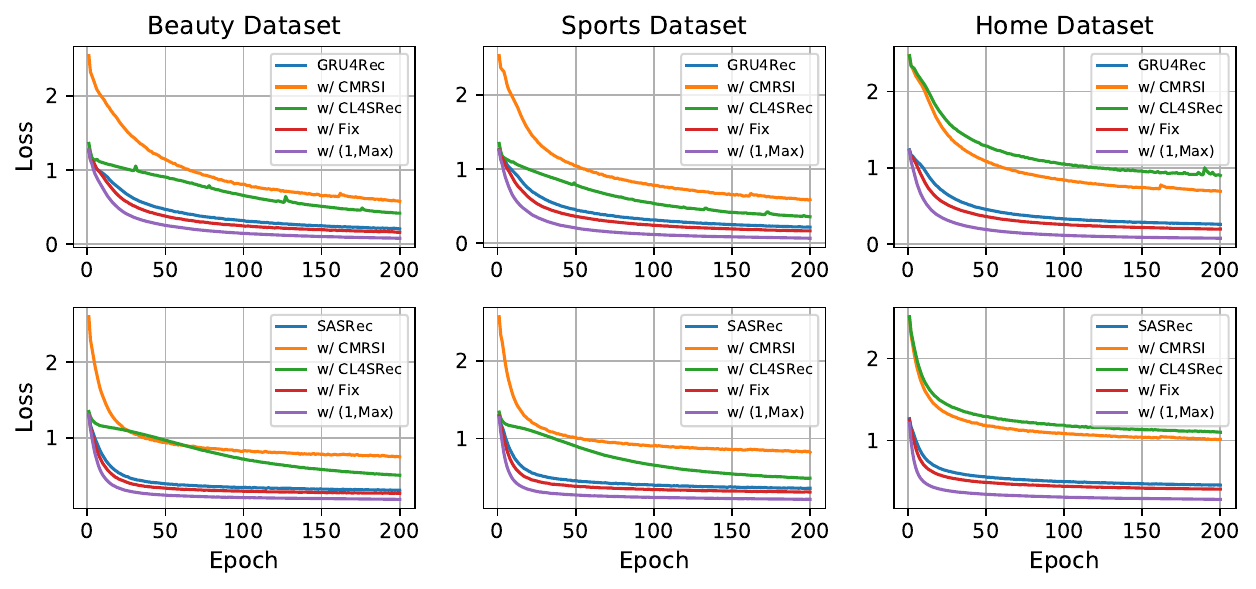}
	\caption{Training loss curve for different methods.}
	\label{fig:loss}
\end{figure}

\section{Why Our Method Works}
The experimental results in Section \ref{sec:Experiments} show that our approach achieves improvements on various types of sequential models, including CNNs, RNNs, MLPs, and Transformers. It is difficult to get a generalized interpretation if we analyze from only a single model type. Therefore, we seek more generalized explanations regarding the common points in all the sequential models.

\subsection{Training Efficiency}
In Figure \ref{fig:loss}, we illustrate the loss convergence of different baseline methods and our RepPad+ during training. We also compare the epoch time consumption and total training time of different methods in Table \ref{tab:speed}. For a fair and intuitive comparison, we only record the time spent on the model training step. The time spent on the validation and eventual testing steps required for the early-stopping strategy is not included.

From Figure \ref{fig:loss}, we can observe that our methods' losses converge slightly more stable and faster than the original model. For CMRSI and CL4SRec, the heuristic data augmentation introduces more training samples along with more noise, resulting in more epochs for the model to converge. The contrastive learning task of CL4SRec also imposes a convergence burden. Compared to the original model, Fix shows a slight improvement in convergence speed, while (1,Max) shows a significant improvement in all cases. From Table \ref{tab:speed}, we can observe that the two variants of RepPad+ only add a small amount of epoch time and total training time compared to the original model. The CMSRI and CL4SRec add extra training samples and auxiliary tasks, substantially increasing computational time. On the Home dataset, which has a large number of interactions, the training time increases by up to a factor of 3. While RepPad+ usually only brings 0.2 to 0.3 times extra training time. Comparing Fix and (1, Max), the former can only perform repeated padding a fixed and smaller number of times. The latter randomly chooses between one time and the maximum number of times, which slightly increases the time-consuming but further improves the loss convergence and model performance. In most cases, (1,Max) can achieve a win-win situation regarding training speed and model performance.

\begin{figure}[!t]
	\centering
	\includegraphics[scale=0.55]{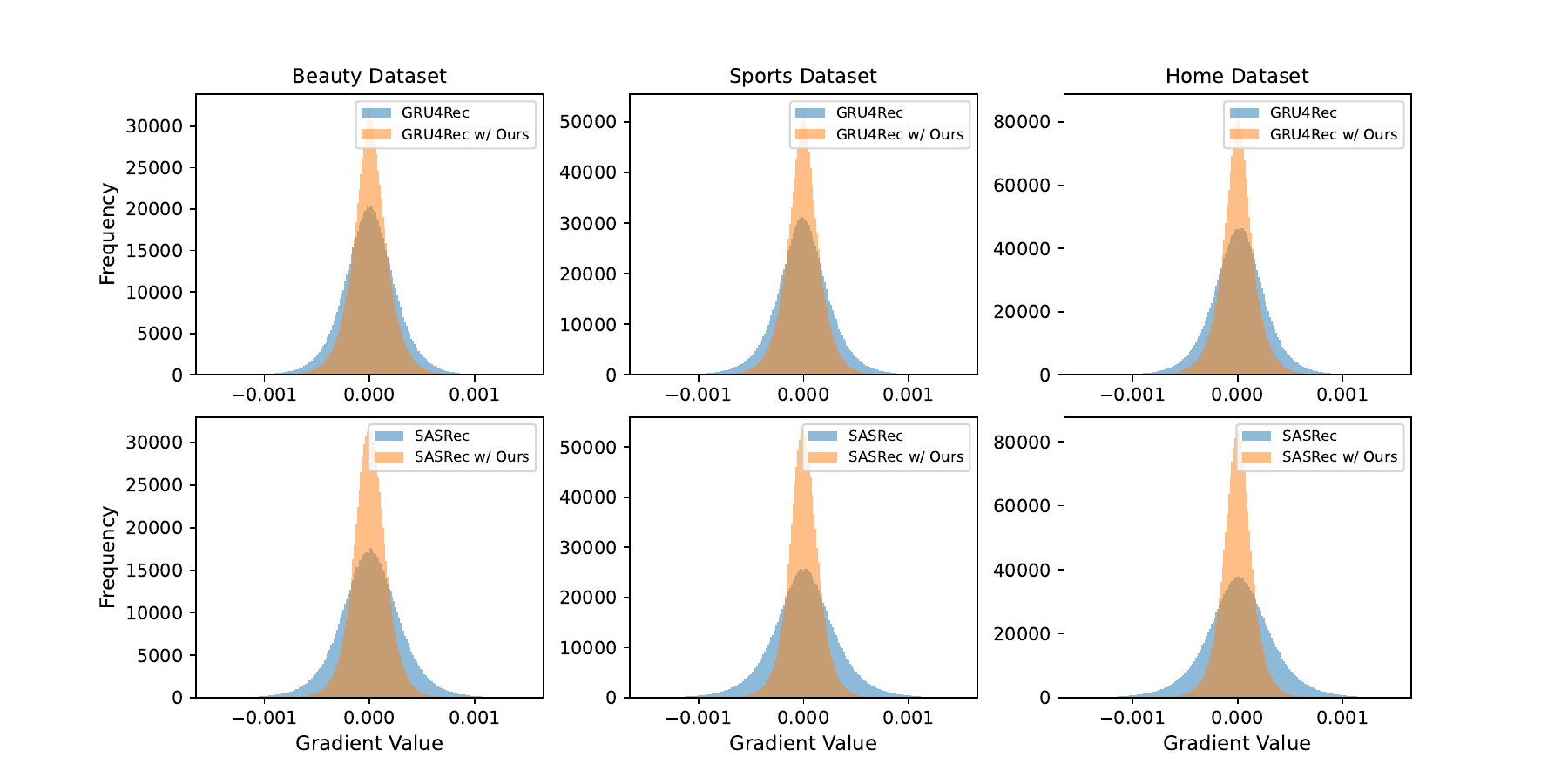}
	\caption{Average gradient value histogram for all item embeddings during training.}
	\label{fig:grad}
\end{figure}

\begin{table}[!t]
  \centering
  \caption{Performance of three special variants of RepPad+.}
\renewcommand\arraystretch{1.0}
    \begin{tabular}{c|c|cccc|cccc}
    \toprule
    \multirow{2}[3]{*}{Dataset} & \multirow{2}[3]{*}{Method} & \multicolumn{4}{c|}{GRU4Rec} & \multicolumn{4}{c}{SASRec} \\
\cmidrule{3-10} & & H@10 & N@10 & H@20 & N@20 & H@10 & N@10 & H@20 & N@20 \\
\midrule
    \multirow{4}[2]{*}{Beauty} & Original & 0.0327 & 0.0162 & 0.0551 & 0.0219 & 0.0637 & 0.0331 & 0.0942 & 0.0408 \\
\cmidrule{2-10} & w/ Ours-$\alpha$ & 0.0312 & 0.0157 & 0.0530 & 0.0197 & 0.0618 & 0.0322 & 0.0903 & 0.0382 \\
         & w/ Ours-$\beta$ & 0.0284 & 0.0144 & 0.0464 & 0.0188 & 0.0622 & 0.0315 & 0.0912 & 0.0394 \\
         & w/ Ours-$\gamma$ & 0.0315 & 0.0158 & 0.0495 & 0.0203 & 0.0533 & 0.0304 & 0.0776 & 0.0365 \\
\midrule
    \multirow{4}[3]{*}{Sports} & Original & 0.0143 & 0.0065 & 0.0249 & 0.0091 & 0.0299 & 0.0160 & 0.0447 & 0.0197 \\
\cmidrule{2-10}         & w/ Ours-$\alpha$ & 0.0127 & 0.0060 & 0.0216 & 0.0081 & 0.0284 & 0.0156 & 0.0413 & 0.0187 \\
         & w/ Ours-$\beta$ & 0.0132 & 0.0061 & 0.0225 & 0.0084 & 0.0289 & 0.0152 & 0.0426 & 0.0184 \\
         & w/ Ours-$\gamma$ & 0.0079 & 0.0038 & 0.0131 & 0.0052 & 0.0198 & 0.0098 & 0.0324 & 0.0130  \\
    \bottomrule
    \end{tabular}%
  \label{tab:variant_2}%
\end{table}%

\subsection{Gradient Stability}
Research has shown that more stable or smooth gradients are more helpful for model training \cite{li2019understanding, reddi2019convergence, xu2019understanding}, thus improving model robustness and performance. In Figure \ref{fig:grad}, we illustrate the average gradient of item embeddings throughout the training process with and without our RepPad+. We show the frequency of the gradient values as histograms. We can observe that the average gradient values are more centralized with the addition of our method. While the gradient is more likely to have extremely large or small values during the original model training, RepPad+ mitigates this phenomenon, making the gradient more stable during the training process. The RepPad+ increases the number of training samples input into the model, especially for short interaction sequences. The model can derive more stable gradient estimates from more samples, thus accurately modeling user preferences.

\subsection{Other forms of RepPad+}
In order to verify whether the effectiveness of RepPad+ is due to increasing the number of iterations in which the data is involved, we implemented RepPad+ in a different form. First, we completed a model training session using (1,Max), during which we recorded the number of times that each sequence was subjected to the execution of RepPad+ and saved it. Afterward, we retrain a model using traditional padding. However, the number of times each sequence is fed into the model at each epoch is determined by the number of times previously recorded. This variant is denoted by Ours-$\alpha$. We also use RepPad+ in the inference phase of the model instead of the training phase. This variant is denoted by Ours-$\beta$. In addition, we replace the padding content of repeated padding with other user sequences. This variant is denoted by Ours-$\gamma$.

The results are illustrated in Table \ref{tab:variant_2}. We can observe that neither using RepPad+ in the inference phase nor changing the number of input data improves the performance of the original model and even results in performance degradation. For Ours-$\alpha$, it treats each repetition of padding as feeding the original sequence into the model once, which changes our RepPad+'s idea of data augmentation from horizontal to vertical but produces a performance degradation compared to the original model. This proves that the use of idle input space impacts the model's training process, which can be seen as adding a regularization to the training process. With this regularization, the model can more equitably focus on short and long sequences in the dataset, providing more accurate recommendations for users with short sequences and reducing the negative impact of long-tail distributions. For Ours-$\beta$, the change in the inference phase will not bring any new knowledge to the model. Given that the training data is distributed with the test data, using RepPad+ at this point will affect the data distribution in the inference phase, which in turn will jeopardize the model performance. From the performance of Ours-$\gamma$, we can observe that randomly padding other sequences leads to a significant degradation of the recommendation performance. This randomized padding strategy interferes with the model's learning of sequence patterns contained in the original data, and the mixing of different sequences may cause the model to learn the incorrect user preferences. In our methods, padding the original sequence maximizes the preservation of that user's preference patterns and avoids the above situation.

\begin{figure}[!t]
	\centering
	\includegraphics[scale=0.37]{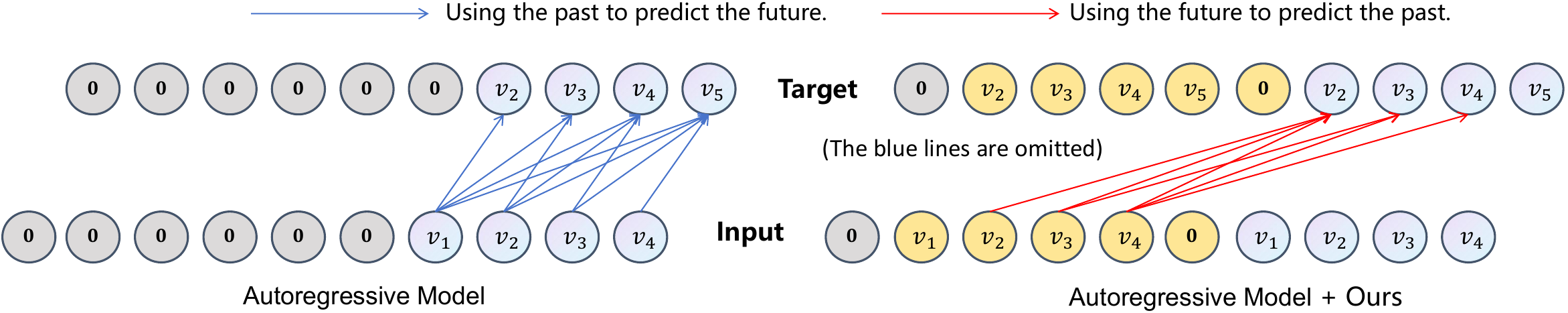}
	\caption{An illustration of information leakage problems for our methods.}
	\label{fig:future}
\end{figure}

\begin{table}[!t]
  \centering
  \caption{The results of future informative leakage analysis. The backbone network is SASRec.}
    \begin{tabular}{c|c|cccc}
    \toprule
    Dataset & Method & H@10 & N@10 & H@20 & N@20 \\
    \midrule
    \multirow{3}[4]{*}{Beauty} & Original & 0.0644 & 0.0335 & 0.0961 & 0.0415 \\
\cmidrule{2-6} & w/ RepPad+ & 0.0762 & 0.0424  & 0.1089 & 0.0507 \\
          & w/ RepPad+ and Avoid Leakage & 0.0699 & 0.0385 & 0.1009 & 0.0467 \\
    \midrule
    \multirow{3}[4]{*}{Sports} & Original & 0.0299 & 0.0160 & 0.0447 & 0.0197 \\
\cmidrule{2-6} & w/ RepPad+ & 0.0369 & 0.0196 & 0.0553 & 0.0242  \\
          & w/ RepPad+ and Avoid Leakage & 0.0346 & 0.0180 & 0.0523 & 0.0225 \\
    \bottomrule
    \end{tabular}%
  \label{tab:future}%
\end{table}%

\subsection{The Leakage of Future Information}
Most sequential recommendation models can be viewed as autoregressive models. They predict the next interaction based on the user's historical sequence data. When training or testing autoregressive models, it is usually necessary to avoid leakage of future information, i.e., the model sees in advance what needs to be predicted when making a prediction. Usually, future information leakage during training can lead to a decrease in the performance and reliability of the model. However, as shown in Figure \ref{fig:future}, we observe that after using our method, there is a leakage of future information, i.e., the model would use future interactions to predict past interactions during training. Also, adding the delimiter \emph{0} does not prevent "predicting the tail with the head". It can be considered as attenuation of attention or influence because the items in the sequence are relatively farther apart after adding the delimiter \emph{0}. 

Counterintuitively, the above information leakage does not degrade the model performance. It may bring more performance gains to the model than the negative impacts it causes. We mask the red line in Figure \ref{fig:future} during training, and the results are presented in Table \ref{tab:future}. We can observe that the model performance decreases after blocking the effect of future information leakage, which proves to be beneficial for recommendation accuracy. Some researchers have also explored using future information to improve model performance \cite{yuan2020future}. In sequential recommendation, although user behavior is in the form of sequential data, sequential dependencies may not be strictly maintained. For example, after a user purchases a cell phone, she may purchase a protective case, headphones, or a screen protector, but there is no sequential dependency between these three items. In other words, the user will likely click on these three items in any order. Our method weakens the notion of strict sequence order in previous sequential recommendations.

\subsection{Summary and Insights}
Based on the analysis above, we believe that our method utilizes the idle input space to provide more samples for model training and improve the training efficiency. Our method can make the gradient update of the model more stable without increasing the total number of training sequences. Unlike existing methods, RepPad augments short sequences to a length comparable to that of long sequences in a nearly lossless manner. RepPad+ improves RepPad by augmenting medium-length sequences, further improving the efficiency of idle space utilization. Existing methods inevitably generate noise or compromise the integrity of the original data when performing augmentation. From another perspective, our method enables the model to focus fairly on short and long sequences, giving more attention to long-tail items and users. In this way, the model can better learn the representation of cold-start users and items and thus give more accurate recommendations. This is similar to L2Aug \cite{wang2022learning}, which categorizes users into core and casual users. The core users tend to have more interaction data, while casual users tend to have short interaction sequences. It uses the historical data from the former to augment the interaction sequence of the latter. However, our method does not include any training process and parameters and does not increase the total amount of training data. Furthermore, utilizing future information is also an interesting finding, similar to the idea presented in GRec \cite{yuan2020future} that future information provides valuable signals about user preferences and can be used to improve the quality of recommendations. How to utilize them wisely is also a valuable research direction. We also speculate that RepPad and RepPad+ act as some kind of regularization during model training, which mitigates the negative effects of the uneven distribution of the original dataset.

\section{Conclusion}
In this work, we argue that the existing sequential recommendation's padding strategy results in a large amount of wasted input space, which can be further utilized to improve the model performance and training efficiency. Based on this idea, we propose an extremely simple yet effective data augmentation method, RepPad+, which leverages the original sequence as padding content instead of the special value \emph{0}. Our method contains no hyper-parameters or learnable parameters compared to existing data augmentation methods. It is a data augmentation plugin with high applicability and generalization. Extensive experiments on multiple datasets demonstrate the superiority of our method. It improves the recommendation performance of various sequential models and outperforms existing data augmentation methods. In addition to recommending performance leadership, our approach is highly competitive in terms of efficiency. We also provide multiple analyses and perspectives into what makes RepPad+ effective. For future work, we plan to explore more effective padding strategies for sequential recommendation. We are also interested in combining RepPad+ with other data augmentation methods.

\section*{Acknowledgments}
This work is partially supported by the National Natural Science Foundation of China under Grant No. 62032013, No. 62102074, and the Science and technology projects in Liaoning Province (No. 2023JH3/10200005).


\bibliographystyle{ACM-Reference-Format}
\bibliography{references}


\end{document}